\documentclass[%
 reprint,
 amsmath,amssymb,
 aps,
]{revtex4-2}

\usepackage{graphicx}
\usepackage{subcaption} 
\usepackage{float} 
\usepackage{dcolumn}
\usepackage{bm}
\usepackage{}

\usepackage{xcolor}

\newcommand{\etaInf}{\eta^{\bowtie}}

\newcommand{\rhoMean}{\langle \rho \rangle}
\newcommand{\mMean}{\langle m \rangle}

\begin{document}

\preprint{APS/123-QED}

\title{Thermodynamic efficiency of self-organisation in nonequilibrium steady states}%

\author{Qianyang Chen}
    \email{Corresponding author: qianyang.chen@sydney.edu.au}
\author{Mikhail Prokopenko}%
 
\affiliation{%
 Centre for Complex Systems, Faculty of Engineering, The University of Sydney, Sydney, NSW 2006, Australia
}%


\begin{abstract}
Active matter generates order or patterns through nonequilibrium dynamics. An open research challenge is to determine how efficiently a nonequilibrium self-organising system can convert consumed energy into macroscopic order.
We study an information-theoretic quantity that directly addresses this challenge by estimating the entropy reduction induced by a small control-parameter perturbation, relative to the generalised work required for the perturbation. This quantity has previously been considered mainly in an equilibrium or near-equilibrium context, and here we extend this framework and apply it to two nonequilibrium self-organising systems: persistent and active Ising models. We observe that the thermodynamic efficiency of nonequilibrium systems maximises at phase transitions, as in equilibrium systems. Furthermore, we compare thermodynamic efficiency and inferential efficiency across control parameters. While these two quantities are equal in equilibrium as a consequence of the fluctuation-dissipation theorem, we report that they diverge out of equilibrium, and the gap serves as a phenomenological signature of broken detailed balance. 
\end{abstract}

\maketitle

\section{Introduction}

Living systems are active and self-organising: they operate far from equilibrium, constantly consume energy, and exchange matter with their environment to establish order and generate functional architectures. Examples of self-organisation in biological systems include flow generation in active microtubule networks \cite{schaller2010Polar, sanchez2012Spontaneous}, morphogenesis in embryo development \cite{deglincerti2016Selforganization, shahbazi2019Selforganization}, collective motion of bacteria, fish and birds \cite{niwa1994self, vicsek2012Collective, cavagna2010Scalefree}. Understanding how such macroscopic structures and coherent functionalities emerge from chaotic, energy-consuming microscopic interactions remains one of the main challenges in biophysics \cite{physoflife}. 

The field of active matter provides fundamental models for understanding the nonequilibrium dynamics of internally-driven collective systems. Examples include active Brownian particles, which model self-propelled motion \cite{shiraishi2023Introduction}; the active vertex model, which describes tissue shape changes arising from the motion of densely packed cells \cite{barton2017Active}; the active gel, which captures forces-exerting active materials, resulting in phenomena such as spontaneous flows, active turbulence, and material buckling~\cite{julicher2018Hydrodynamic}; and large-scale active systems capable of motility-induced phase separation and polar alignment~\cite{crosato2019irreversibility}. Understanding what governs the efficiency of an active system converting consumed energy into structured order is an ongoing problem.

Much of the existing literature on efficiency focuses on the conversion of input energy into mechanical work, known as \textit{thermal efficiency}. Stochastic thermodynamics extends this macroscopic notion to smaller systems that are strongly influenced by thermal fluctuations \cite{vandenbroeck2005Thermodynamic, benenti2011Thermodynamic, brandner2013Strong, pietzonka2018Universal}, and refers to it as \textit{thermodynamic efficiency}. A classic example is a molecular motor, which consumes chemical energy provided by ATP to generate mechanical motion \cite{vale2003Molecular, seifert2025Stochastic}. Closely related energy-based definitions are also used in biological contexts \cite{westerhoff1983Thermodynamic, kempes2017Thermodynamic}.

In this study, however, we use the notion of thermodynamic efficiency drawing from an information-theoretic framework. In this setting, given the change in the control parameter, the thermodynamic efficiency quantifies the conversion of work into gained predictability (i.e., reduction in Shannon entropy) \cite{crosato2018Thermodynamics}. Simulation studies have examined this ratio of predictability gain to work for various systems, including canonical magnetisation models \cite{nigmatullin2021Thermodynamic, chen2025Why, chen2026Generalising}, self-propelled particles \cite{crosato2018Thermodynamics}, urban development \cite{crosato2018Critical}, and contagion dynamics \cite{harding2018Thermodynamic}. However, these studies assume dynamics at or near equilibrium. Extension of this analysis to nonequilibrium dynamics remains an open challenge.

We note that there are other information-theoretic definitions of thermodynamic efficiency. For example, Allahverdyan et al. \cite{allahverdyan2009Thermodynamic} define thermodynamic efficiency as the ratio of change in mutual information to entropy production for an interacting Brownian particle model. More broadly, Peliti and Pigolotti \cite[sec. 2.2]{peliti2021Stochastic} extended the concept of thermal efficiency from energy-driven heat engines to information-driven mesoscopic engines, defining thermodynamic efficiency as the ratio of input entropy over output entropy. We emphasise that throughout this paper, thermodynamic efficiency refers specifically to entropy reduction over work, given the change in the corresponding control parameter.

How efficiently can a nonequilibrium self-organising system convert consumed energy into macroscopic order, and how does this energy-to-order efficiency compare to an equilibrium counterpart? Here, we restrict our study to collective systems in nonequilibrium steady states (NESS), where the external drive of the system is constant, and the probability distribution over system states remains stationary \cite{seifert2012Stochastic, peliti2021Stochastic, shiraishi2023Introduction, seifert2025Stochastic}. By examining the thermodynamic efficiency of a NESS system across parameter space, we aim to understand how changes in the control parameter influence the coordination of the collective system in a nonequilibrium setting.

We studied two spin-lattice models that use different driving mechanisms. The first is the persistent Ising model of Kumar and Dasgupta \cite{kumar2020Nonequilibrium}, which deviates only minimally from the equilibrium Ising model: nonequilibrium is introduced through a modified spin-flip rule, in which a constant drive biases the usual Metropolis or Glauber acceptance probability, thereby breaking detailed balance. The second is the active Ising model introduced by Solon and Tailleur \cite{solon2013Revisiting, solon2015Flocking}: a system of spin-carrying particles that both align ferromagnetically through spin flips and hop on the lattice with a spin-dependent bias, and nonequilibrium arises from the self-propelled motion. Together, these two examples allow us to compare two distinct mechanisms to drive a system out of equilibrium: one via modified spin dynamics and the other via active transport.

The paper is organised as follows. Section~\ref{sec:method} reviews the definition of thermodynamic efficiency and describes the two models considered in this study. Section~\ref{sec:results} presents the results for each model. Section~\ref{sec:discussion} discusses their interpretation and broader implications, drawing on both cases to identify more general conclusions.

\section{Methods \label{sec:method}}

\subsection{Thermodynamic efficiency}

Consider a collective system controlled by a set of parameters $\{\lambda_1, \lambda_2, ..., \lambda_n\}$. We perturb the system's control parameter by a small amount $\delta \lambda$, and measure the resulting reduction in Shannon entropy or gain in predictability, $-\delta S$, and the infinitesimal work associated with the perturbation, $\langle\delta W \rangle$, where $\langle .\rangle$ is the ensemble average. The thermodynamic efficiency $\eta$ is the ratio between the entropy reduction and the corresponding infinitesimal work. Formally, for perturbation of the $j^{\text{th}}$ control parameters, the thermodynamic efficiency is defined as \cite{crosato2018Thermodynamics,chen2026Generalising}:
\begin{equation} \label{eq:eta}
    \eta(\lambda_j) = -\frac{\partial S}{\partial \lambda_j}\bigg /\frac{\partial \langle W\rangle}{\partial \lambda_j}.
\end{equation}
Shannon entropy for the probability distribution of the system's microstate $x$ is defined as:
\begin{equation}
    S = -\sum_x p(x;\underline{\lambda})\ln p(x;\underline{\lambda}),
\end{equation}
and work associated with the small perturbation $\delta \lambda$ is obtained by the change in Hamiltonian $\mathcal{H}$ before and after the perturbation, assuming that the system's configuration $\xi$ does not change during the perturbation \cite[Sec 3.2.2]{seifert2025Stochastic}:
\begin{equation}
    W_{\lambda \to \lambda+\delta \lambda} = \mathcal{H}(\xi; \lambda+\delta \lambda) - \mathcal{H}(\xi; \lambda),
\end{equation}
which can then be averaged over the ensemble to obtain $\langle \delta W\rangle$.

At equilibrium, the probability distribution of the system's microstate follows the Gibbs distribution:
\begin{equation}
    p(x;\underline{\lambda}) = \frac{1}{Z}e^{-\sum_i \lambda_i\,X_i(x)},
\end{equation}
where $X_i$ is the observable (or collective variable) conjugate to the control parameter $\lambda_i$ \cite{crooks2007Measuring}. Under the equilibrium assumption, thermodynamic efficiency has been shown to have an equivalent ``inferential" form \cite{chen2026Generalising}:
\begin{equation}
    \eta(\lambda_j) = \etaInf(\lambda_j) = -\frac{\sum_i \lambda_i \text{Cov}(X_i, X_j)}{\langle X_j \rangle}.
\end{equation}

The thermodynamic efficiency $\eta$ provides a system-centric measure of the system's ability to translate work into predictability gain or intrinsic cohesiveness as it self-organises. The inferential form $\etaInf$ offers an observer-centric interpretation of thermodynamic efficiency: at equilibrium, high thermodynamic efficiency means the system's hidden parameters can be inferred from observables with greater accuracy, as the system becomes more sensitive to changes in control parameters. The equivalence of $\eta$ and $\etaInf$ is a direct consequence of the fluctuation-dissipation theorem and can be derived using the Gibbs distribution \cite{chen2026Generalising}. Out of equilibrium, however, neither the Gibbs distribution nor the equality holds. In this study, we examine how $\eta$ diverges from $\etaInf$ as the system is driven out of equilibrium. To distinguish them, we refer to $\etaInf$ as \textit{inferential efficiency} henceforth.

\subsection{Entropy production rate}
The entropy production rate $\dot{S}$ quantifies the irreversibility of a system's dynamics. The ensemble entropy production rate of the combined system and bath is given by \cite{schnakenberg1976Network, vandenbroeck2005Thermodynamic}:
\begin{equation}
    \dot{S}^{tot} = k_B \sum_{x, x'} (P_{x}W_{x \to x'})\ln \frac{P_{x}W_{x \to x'}}{P_{x'}W_{x' \to x}},
\end{equation}
summing over all permutations of states $x, x'$. Here, $W_{x \to x'}$ denotes transition rate from state $x$ to $x'$, and $P_x$ is the probability of the system being in state $x$. The product $P_{x}W_{x \to x'}$ represents the total probability flux from state $x$ to $x'$. The log ratio $\ln \frac{P_{x}W_{x \to x'}}{P_{x'}W_{x' \to x}}$ measures how strongly forward transition is biased relative to the reverse transition. The entropy production rate can be decomposed into two components: the system's entropy change $\dot{S}^{sys}$ and the entropy change of the environment $\dot{S}^{env}$ \cite{vandenbroeck2005Thermodynamic, noa2019Entropy,landi2021Irreversible, aguilera2023Nonequilibrium}:
\begin{equation}
    \dot{S}^{tot} = k_B \underbrace{\sum_{x, x'} (P_{x}W_{x \to x'})\ln \frac{P_{x}}{P_{x'}}}_{\dot{S}^{sys}} + \underbrace{\sum_{x, x'} (P_{x}W_{x \to x'})\ln \frac{W_{x \to x'}}{W_{x' \to x}}}_{\dot{S}^{env}}.
\end{equation}
In NESS, $\dot{S}^{sys} = 0$ as the probability distribution $P$ does not change. The total entropy production rate is equal to the entropy change of the environment $\dot{S}^{env}$, meaning that all the entropy produced by the system is dissipated into the environment \cite{landi2021Irreversible}. The entropy production rate is always non-negative, and is zero only at equilibrium, where detailed balance holds.

\subsection{Persistent Ising model}
We study a ferromagnetic model with a non-conservative drive $E_0$ introduced to the transition dynamics, known as the persistent Ising model \cite{kumar2020Nonequilibrium}. Under the modified update rule, spin configurations persist longer for $E_0>0$. The Hamiltonian of the Ising system in the absence of a magnetic field is given by:
\begin{equation} \label{eq:hamiltonian}
    \mathcal{H(\underline{\sigma})} = -J \sum_{\langle ij \rangle}\sigma_i\sigma_j,
\end{equation}
where $\sigma_i \in \{-1, 1\}$ denotes the spin of the $i^{\text{th}}$ site, and $\sum_{\langle ij \rangle}$ represents a sum over all nearest-neighbour pairs $(i,j)$. The coupling strength $J$ is the control parameter.

The interaction energy of a single spin $\sigma$ with its nearest neighbours $\mathcal{N}(\sigma)$ is:
\begin{equation}
    E(\sigma) = -J \sigma\sum_{j \in \mathcal{N}(\sigma)} \sigma_j.    
\end{equation}

In the persistent Ising model, the transition probability of a single spin-flip $\sigma \rightarrow -\sigma$ is determined by both the energy difference before and after the spin-flip, and the drive $E_0$. The effective energy difference $\Delta \tilde{E}$ is:
\begin{equation} \label{eq:pim_E_tilde}
\begin{split}
\Delta \tilde{E}_{\sigma \rightarrow -\sigma} &= E(-\sigma) - E(\sigma) + E_0 \\
&= 2 J \sigma \sum_{j \in \mathcal{N}(\sigma)} \sigma_j + E_0
\end{split}
\end{equation}

Using Glauber dynamics \cite{glauber1963TimeDependent}, the modified transition probability is:
\begin{equation} \label{eq:pim_glauber}
\omega(\sigma \rightarrow -\sigma) = \frac{1}{1+\exp(\beta \Delta \tilde{E}_{\sigma \rightarrow -\sigma})}.
\end{equation}

\begin{figure}[h]
\centering
\begin{minipage}[c]{0.5\linewidth}
  \centering
  \subcaptionbox{2D square lattice and modified transition rates.}{%
  \includegraphics[width=\linewidth]{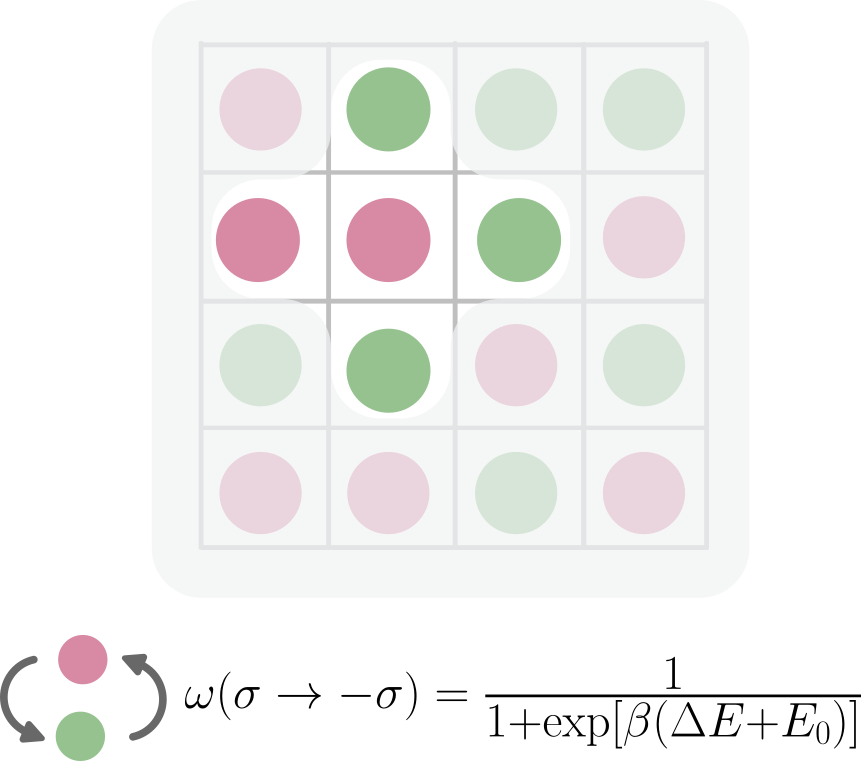}
  }
\end{minipage}
\hfill
\begin{minipage}[c]{0.48\linewidth}
  \centering
  \subcaptionbox{Flips suppressed.}{%
  \includegraphics[width=0.7\linewidth]{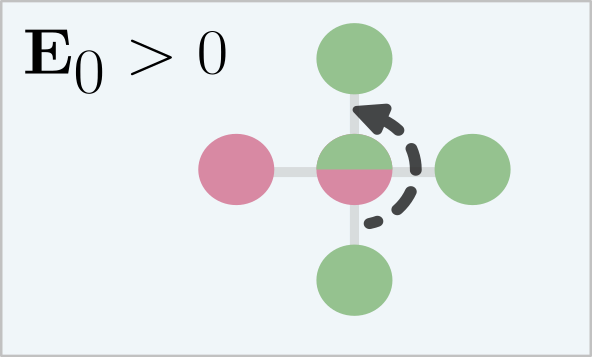}
  }\\[3ex]
  \subcaptionbox{Flips promoted.}{%
  \includegraphics[width=0.7\linewidth]{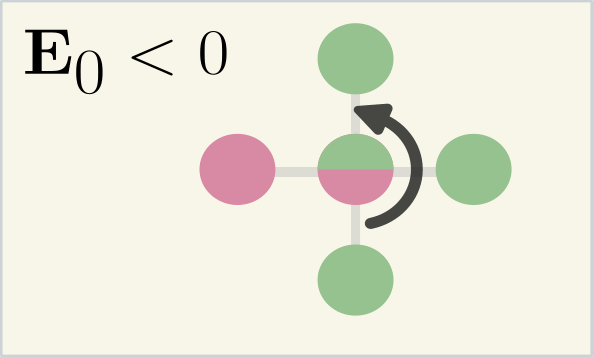}
  }
\end{minipage}

\caption{\textbf{Schematic of persistent Ising model}. (a) We consider the square-lattice Ising model with nearest-neighbour interactions, modified by a non-conservative drive $E_0$ as described in \cite{kumar2020Nonequilibrium}. The transition probability of a single spin is determined by the effective change in the energy $\Delta \tilde{E} = \Delta E + E_0$, where $\Delta E$ is the energy difference before and after the spin-flip. (b) When $E_0>0$, there is a cost associated with the spin-flip. Transitions in all directions are inhibited. (c) When $E_0<0$, the spins are self-excited. Transitions in all directions are promoted.}
\label{fig:PIM_schematic}
\end{figure}

The constant drive $E_0$ can be interpreted as an activation offset, as illustrated in Figure \ref{fig:PIM_schematic}. A positive $E_0$ imposes an additional kinetic barrier or a ``cost" to each spin-flip, thereby suppressing transition in both directions, regardless of whether the flip will reduce alignment or increase alignment. Conversely, a negative $E_0$ uniformly lowers the barrier, promoting the transitions in both directions. A system with $E_0<0$ acts as if the spin transitions are self-excited. Either way, the detailed balance condition is broken when $E_0 \neq 0$, and the system is out of equilibrium. For the rest of this study, we use cold colours (light and dark blue) to indicate cases where the activities are inhibited ($E_0>0$), warm colours (peach, red) to indicate cases where activities are promoted ($E_0<0$), and grey represents the equilibrium ($E_0=0$) baseline.

\subsection{Active Ising model} \label{sec:model_AIM}

We study a second nonequilibrium model: the active Ising model \cite{solon2013Revisiting, solon2015Flocking}, where particle self-propulsion drives the system out of equilibrium. The model consists of N particles hopping on a two-dimensional square lattice of size $L_x \times L_y$ sites with periodic boundary conditions. Each particle carries a spin of $\sigma = \{-1, +1\}$. For site $i$ containing $n_i^+$ positive spins and $n_i^-$ negative spins, its local density is $\rho_i = n_i^+ + n_i^-$, and the local magnetisation is $m_i = n_i^+ - n_i^-$.

\begin{figure}[h]  
    \centering
    \includegraphics[width=0.5\textwidth]{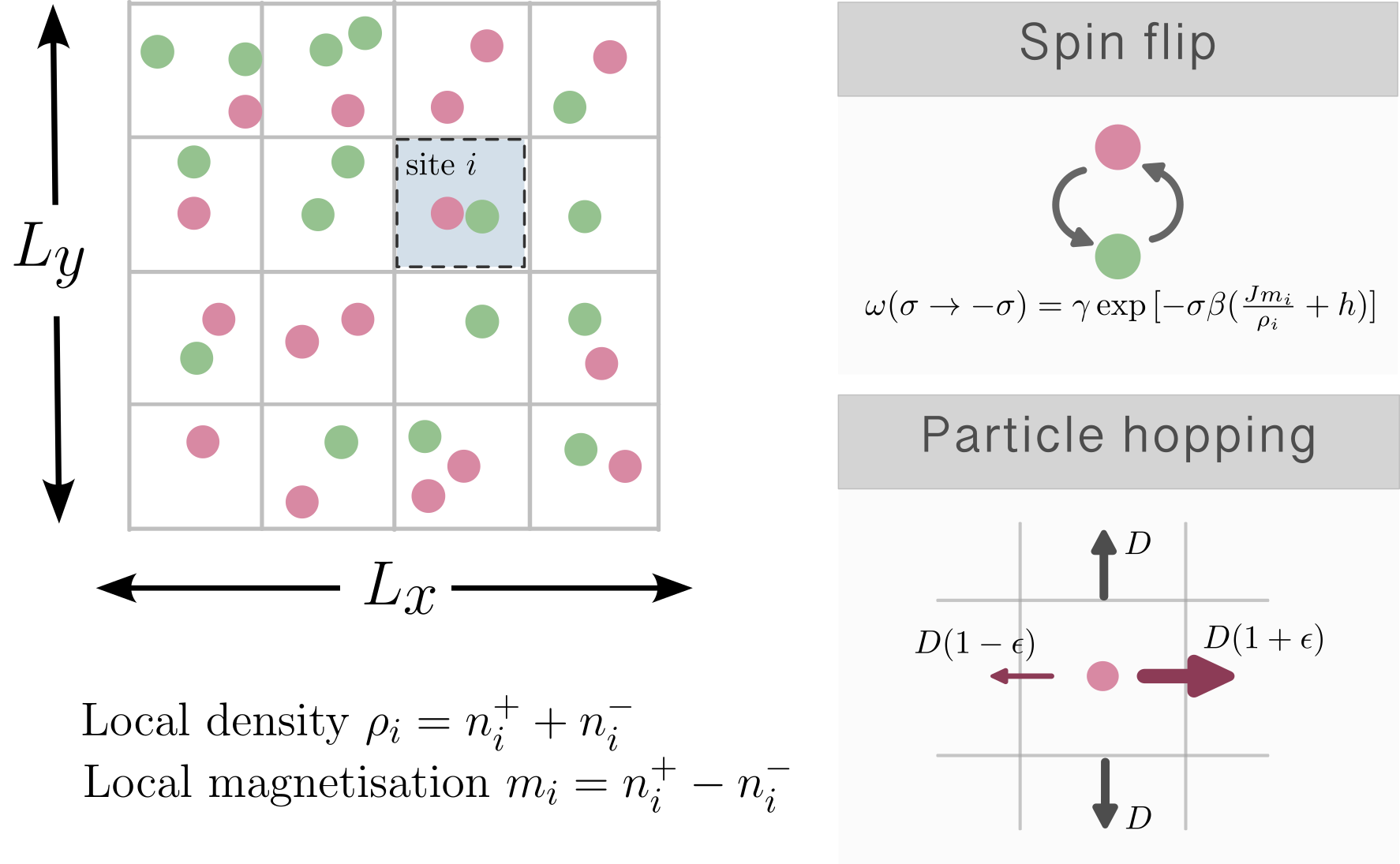}
    \caption{\textbf{Schematic of active Ising model}. The model consists of $N$ spin-carrying and self-propelled particles on a lattice of size $L_x \times L_y$, where each site contains an arbitrary number of particles (left panel). The particles undergo two different dynamics (right panel): spin-flipping and hopping. Local density $\rho_i$ and local magnetisation $m_i$ are measured for each site.}
    \label{fig:AIM_schematic}
\end{figure}

Each particle undergoes two types of dynamics: lattice hopping and spin flipping (Figure \ref{fig:AIM_schematic}). A particle hops vertically with a symmetric rate $D$, whereas the horizontal motion is spin-dependent, with rate $D(1-\varepsilon\sigma)$ to the left and $D(1+\varepsilon\sigma)$ to the right. $\varepsilon$ controls the degree of self-propulsion, where $\varepsilon=0$ gives purely diffusive motion and $\varepsilon=1$ for fully self-propelled motion. 

For the spin-flip dynamics, we extend the model to include the coupling strength $J$ between interacting particles and the external magnetic field $h$. This extension allows the change of control parameter to be described by the Hamiltonian dynamics and enables us to compute the associated work. The Hamiltonian is:
\begin{equation}
    \mathcal{H}(\underline{\rho}, \underline{m}) = \sum_i [ -J(\frac{m_i^2}{2 \rho_i} - \frac{1}{2}) - hm_i].
\end{equation}

The spin-flip rate given the modification is therefore:
\begin{equation} \label{eq:AIM_spin_flip}
    \omega(\sigma \to -\sigma) = \gamma \exp[-\sigma\beta(\frac{Jm_i}{\rho_i} + h)],
\end{equation}
where $\beta$ is the inverse temperature and the rate $\gamma$ is taken to 1 without loss of generality.

In certain parameter regimes, the system exhibits phase separation as a polarised band travelling on a homogeneous background, making it a useful minimal model for flocking behaviour. In this study, we use the coupling strength $J$ and the external magnetic field $h$ as control parameters. We treat the average density $\rho_0 = \frac{N}{L_xL_y}$, temperature $T$ (or inverse temperature $\beta=\frac{1}{k_BT}$, where $k_B$ is the Boltzmann constant), the degree of self-propulsion $\varepsilon$ and diffusion rate $D$ as hyperparameters. 

The self-propulsion speed ranges from $\varepsilon=0$ (purely diffusive) to $\varepsilon=1$ (fully self-propelled). Even at $\varepsilon=0$, the system remains out of equilibrium due to the diffusive motion \cite{solon2015Flocking}. At fixed temperature, varying average density $\rho_0$ and self-propulsion speed $\varepsilon$ changes only the width of the polarised band, not the general behaviour of the system \cite{solon2015Flocking}. Changing temperature $T$ controls thermal noise in the spin-flip dynamics and has the same effect as varying the coupling strength $J$. The diffusion rate $D$ sets the timescale for particle hopping relative to alignment dynamics. In the main sections of this study, we set $\rho_0 = 3$, $\beta = 1$, $\varepsilon=0.9$ and $D = 1$ for all simulations. We also explored different values of particle density $\rho_0$ and self-propulsion speed $\varepsilon$ to test robustness of the results (see Appendix \ref{ap:AIM_simulation}).

\section{Results \label{sec:results}}

\subsection{Persistent Ising model}
The system is in equilibrium when the drive $E_0=0$, and nonequilibrium steady state (NESS) when $E_0\neq0$. Using the coupling strength $J$ as the control parameter, we identify the average spin-spin interaction, $\langle \sum_{ij} \sigma_i \sigma_j/N \rangle$, as the conjugate order parameter. We note that the original study by Kumar and Dasgupta \cite{kumar2020Nonequilibrium} considered temperature $T$ as the control parameter and average magnetisation $|m|$ as the order parameter. We use $J$ instead of $T$ as the control parameter to separate the effects of parameter tuning from those of the nonequilibrium drive $E_0$.

\begin{figure}[h]
    \begin{subfigure}[t]{0.35\textwidth}
        \includegraphics[width=\textwidth]{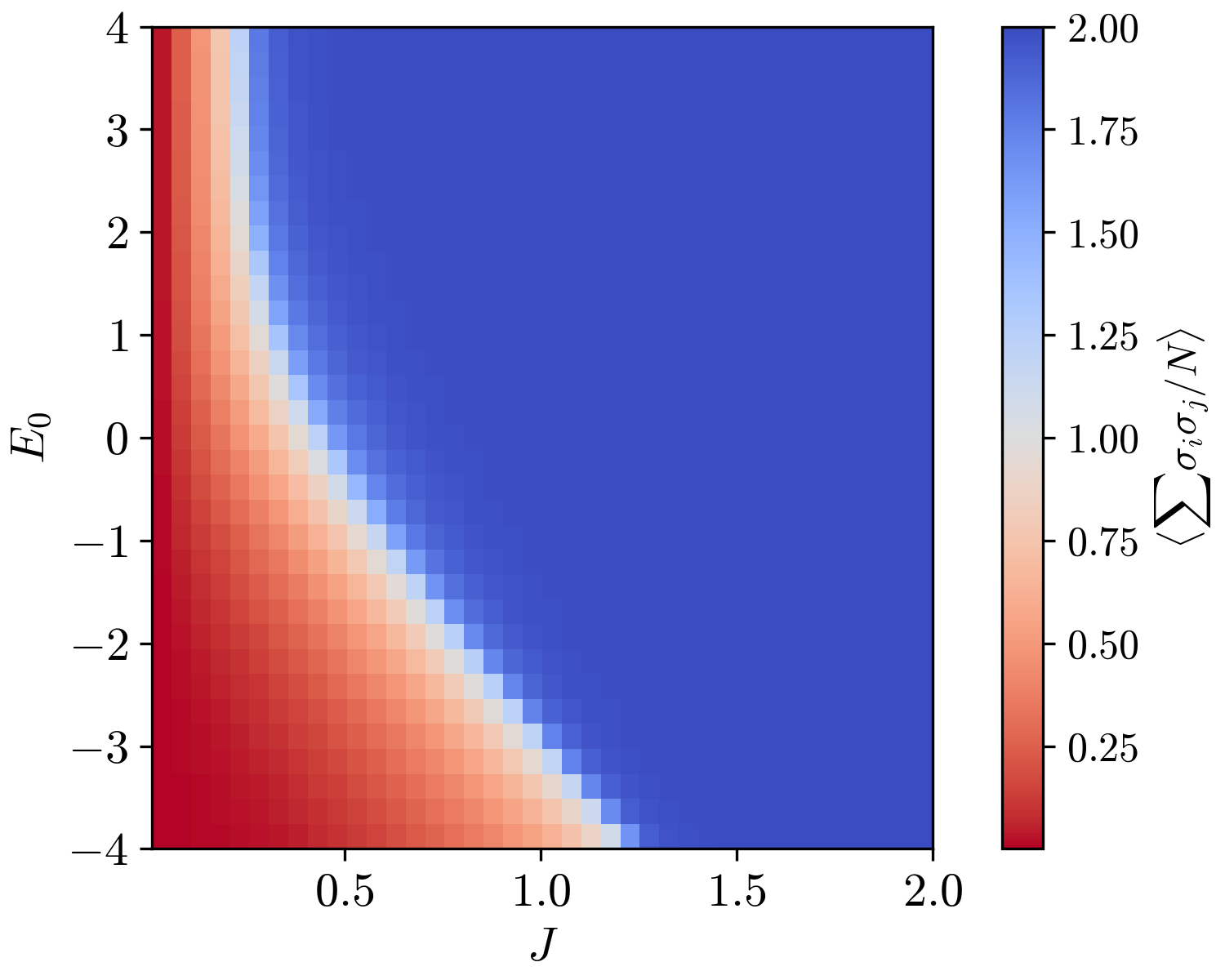}
        \caption{Phase plot of the system by varying control parameter $J$ and drive $E_0$}
        \label{fig:PIM_phase_plot}
    \end{subfigure}  
    \vfill
    \begin{subfigure}[t]{0.33\textwidth}
        \includegraphics[width=\textwidth]{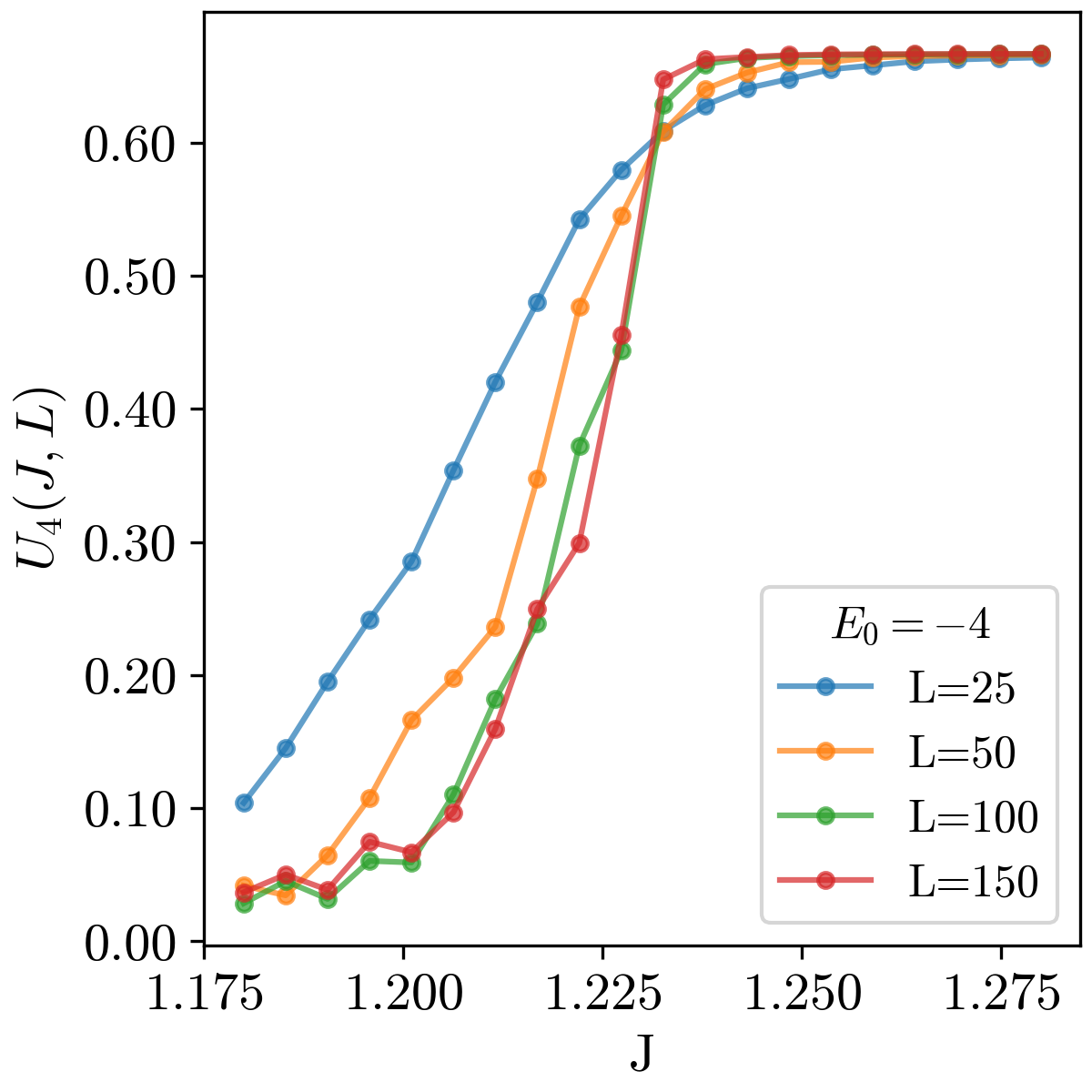}
        \caption{Binder cumulant $U_4(J, L)$ for $E_0=-4$}
        \label{fig:PIM_binder_cumulant}
    \end{subfigure}
    
    \caption{\textbf{Phase transition and critical regime in the persistent Ising model}. \textbf{(a)} The conjugate order parameter is shown as a function of control parameter $J$ and drive $E_0$. The curve separating two phases of the system represents the critical regime. Lattice size: $50 \times 50$. \textbf{(b)} Binder cumulant $U_4(J, L)$ for $E_0=-4$. The crossing point for different lattice sizes yields the critical coupling strength $J_c(E_0=-4) \approx 1.2315$.}
    \label{fig:phase_transition}
\end{figure}

In both equilibrium and NESS, the system undergoes a phase transition when the control parameter $J$ crosses a critical value. Figure \ref{fig:phase_transition}(a) shows the phase plot on the $J-E_0$ plane. The critical regime divides the system into two regimes: a strongly-coupled (i.e., ordered) phase, where the neighbouring spins tend to align and average interaction is high (blue), and a weakly-coupled (i.e., disordered) phase, where the average interaction is close to zero (red). Figure \ref{fig:phase_transition}(b) shows the Binder cumulant $U_4 = 1 - \langle m^4 \rangle / 3 \langle m^2 \rangle^2$ for a specific value of $E_0=-4$. We estimate $J_c$ from the crossing points of the Binder cumulant for different lattice sizes, and the critical $J$ for different $E_0$ are summarised in Table \ref{tab:critical_J}. 
\begin{table}[b]
    \caption{\label{tab:critical_J} Critical coupling strength $J_c$ for different drives $E_0$. The critical coupling strength $J_c$ is numerically identified using crossing points of Binder cumulant for lattice sizes $L = 25, 50, 100, 150$.}
    \begin{ruledtabular}
    \begin{tabular}{ccc}
    \textrm{Drive $E_0$}&
    \textrm{$J_c$}&
    \textrm{}\\
    \colrule
    -4 & 1.2315 &  (numerical)\\
    -2 & 0.8026 &  (numerical)\\
    0  & 0.4395 & (known theoretical result 0.4407) \\
    2  & 0.2603 &  (numerical)\\
    4  & 0.2234 &  (numerical)\\
    \end{tabular}
    \end{ruledtabular}
\end{table}
Note that when $E_0=0$, the model reduces to a two-dimensional Ising model with zero external magnetic field. At the critical point, the coupling strength $J$ and temperature $T$ therefore satisfy $\frac{k_B T}{J} = \frac{2}{\ln(1+\sqrt{2})}$ \cite{onsager1944Crystal}, which gives $J_c(E_0=0) \approx 0.4407$ under the convention $k_B T = 1$. For $E_0 \neq 0$, no analogous analytical relation is known between $J_c(E_0)$ and $T_c(E_0)$. Thus, $J_c$ cannot be computed directly from the critical temperatures reported in \cite{kumar2020Nonequilibrium}.

\begin{figure}[h] 
    \centering
    \includegraphics[width=0.33\textwidth]{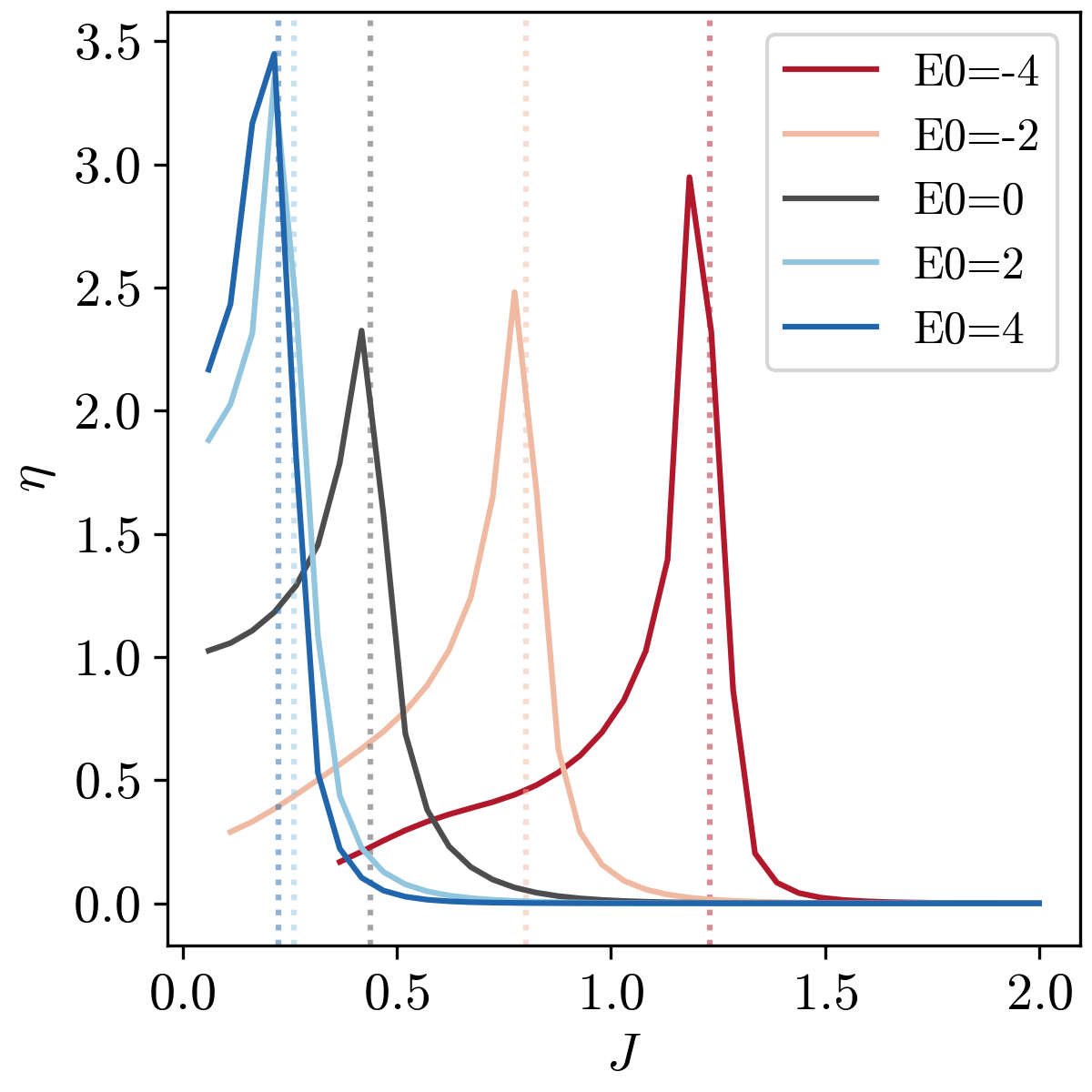}
    \caption{\textbf{Thermodynamic efficiency of persistent Ising model}. Thermodynamic efficiencies $\eta(J)$ plotted against the control parameter $J$, at different external drives $E_0$. Lattice size: $100 \times 100$.}
    \label{fig:eta}
\end{figure}

Figure \ref{fig:eta} shows thermodynamic efficiency computed for different $E_0$, with vertical dotted lines indicating the corresponding critical values $J_c$. In the equilibrium case ($E_0=0$), thermodynamic efficiency $\eta$ peaks at the critical point, consistent with the earlier studies \cite{nigmatullin2021Thermodynamic, chen2025Why, chen2026Generalising}. Remarkably, $\eta$ also peaks near the corresponding $J_c$ in NESS, despite small numerical deviations. This suggests that the thermodynamic efficiency of the collective system is maximised in the critical regime, even for nonequilibrium cases.

\begin{figure}[h]
    \centering
    \includegraphics[width=0.48\textwidth]{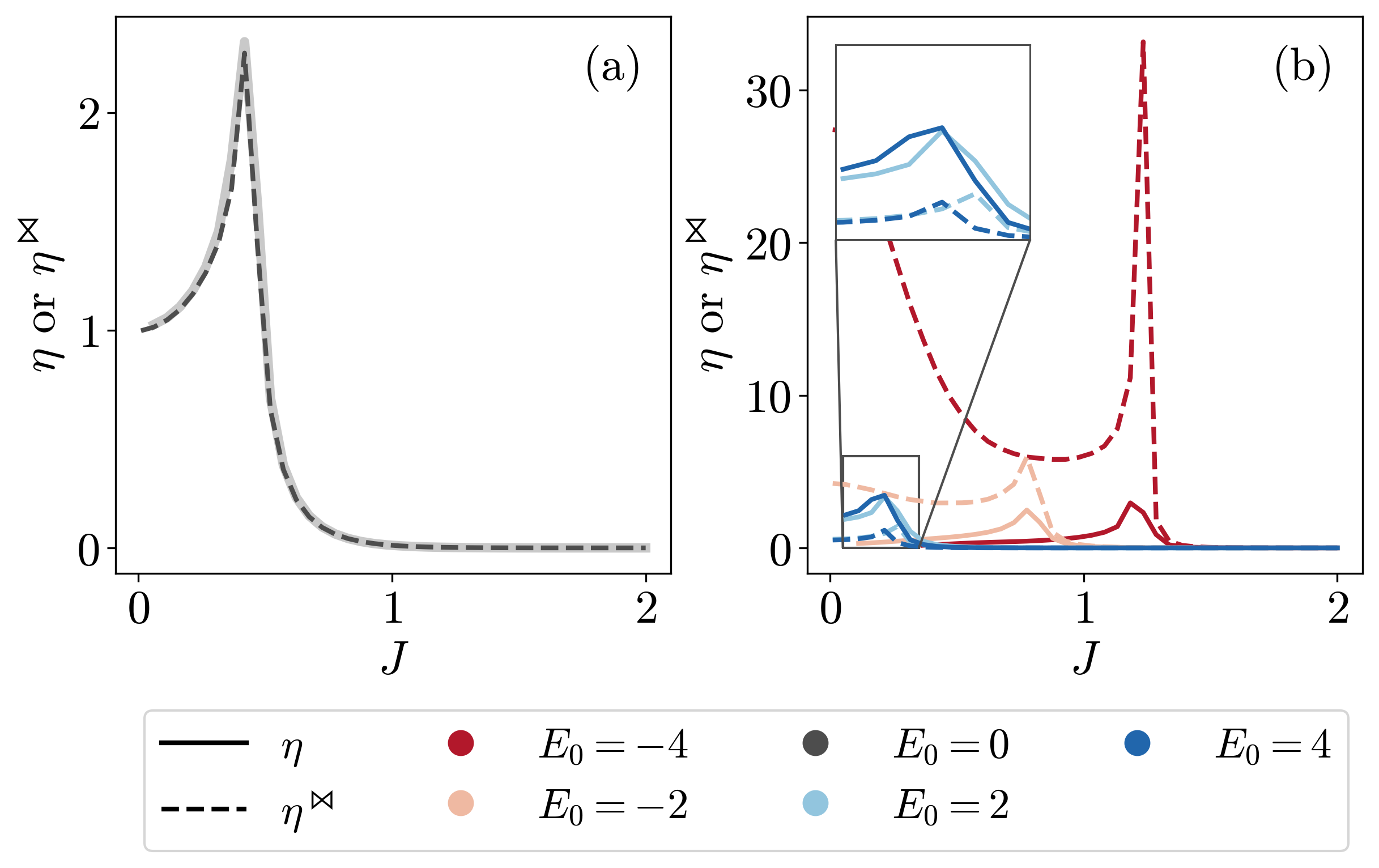}
    \caption{\textbf{Thermodynamic efficiency ($\eta$) compared with inferential efficiency ($\etaInf$)}. Lattice size: $100 \times 100$.}
    \label{fig:pim_compare_eta}
\end{figure}

Next we compare the thermodynamic efficiency and inferential efficiency. Figure~\ref{fig:pim_compare_eta}(a) shows the equilibrium baseline, where $\eta(J) = \etaInf(J)$. In NESS (Figure~\ref{fig:pim_compare_eta}(b)), the peaks of $\eta(J)$ and $\etaInf(J)$ still align, indicating that the system achieves optimal thermodynamic and inferential efficiencies within the same control-parameter regime. Their magnitudes, however, no longer coincide. When $E_0>0$ (light and dark blue curves), we find that $\eta(J) > \etaInf(J)$. In this regime, the spin-flip activities are suppressed, which reduces the fluctuation of the macroscopic observable and lowers $\etaInf(J)$ relative to $\eta(J)$. On the other hand, when $E_0<0$, self-excited spin-flips lead to higher activities, thus increasing inferential efficiency. However, these drive-induced fluctuations do not translate into stable structural order as captured by the thermodynamic efficiency.

\begin{figure}[h]
    \centering
    \includegraphics[width=0.33\textwidth]{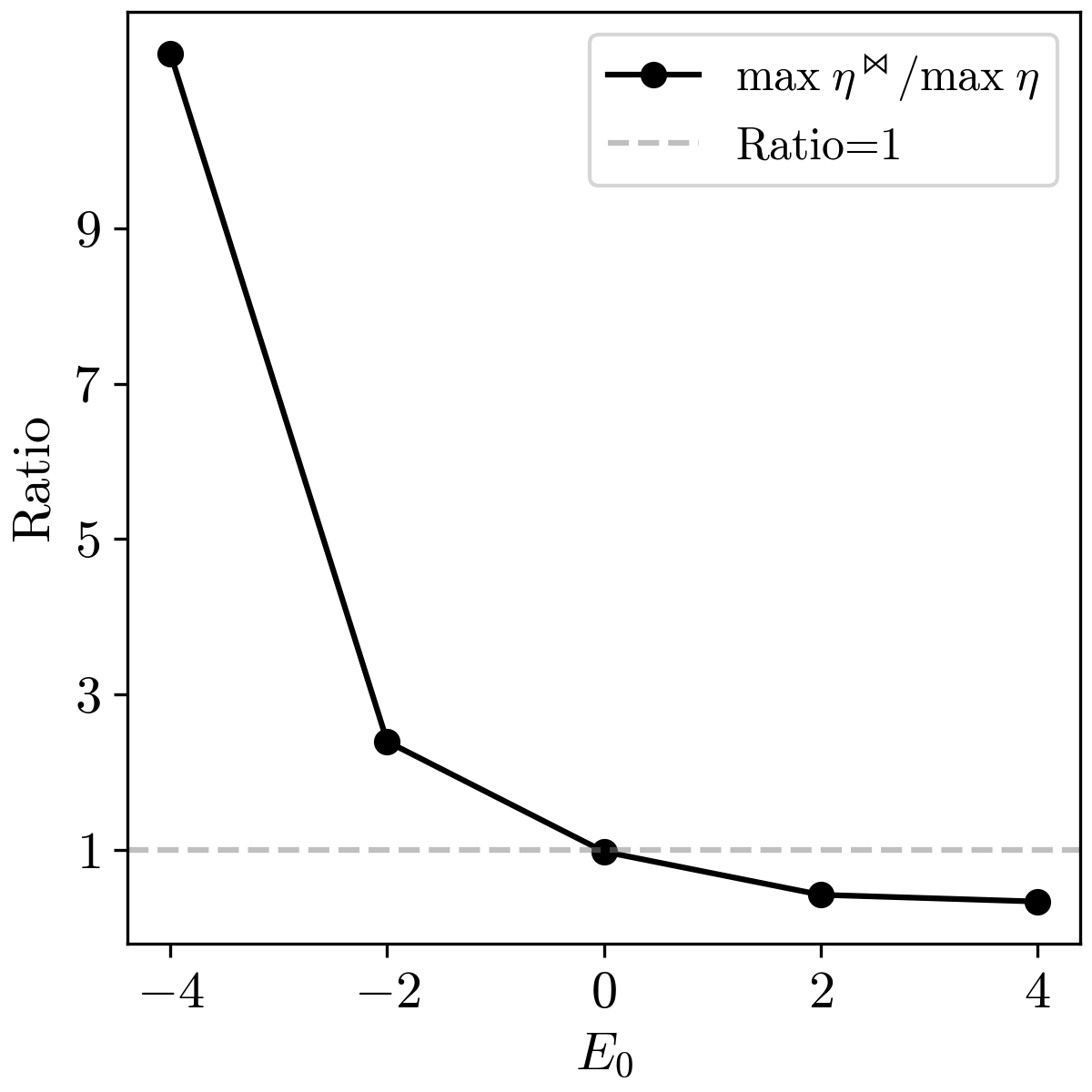}
    \caption{\textbf{Ratio of inferential efficiency ($\etaInf$) and thermodynamic efficiency ($\eta$) plotted against different $E_0$}. Lattice size: $100 \times 100$.}
    \label{fig:eta_ratio}
\end{figure}

The divergence between the maxima of $\etaInf$ and $\eta$ provides a phenomenological signature of broken detailed balance and the associated violation of the fluctuation-response relation. We track the ratio of these maxima against different values of nonequilibrium drive $E_0$ (Figure~\ref{fig:eta_ratio}). A ratio $>1$ indicates that the system exhibits stronger fluctuations than would be expected from the self-organising process, which is the case for $E_0<0$ (i.e., promoted activities). The ratio is less than 1 for $E_0>0$ (i.e., suppressed activities), and exactly 1 when the system is at equilibrium. Moreover, the ratio scales asymmetrically with positive and negative drive. To give some intuitions, we can compare $E_0 = 4$ with $E_0 = -4$. The ratio between the spin transition rate and its reciprocal can be rewritten as:
\begin{equation}
    \frac{\omega(\sigma \to -\sigma)}{\omega(-\sigma \to \sigma)} = \frac{\exp(-\beta E_0) + \exp(-\beta \Delta E)}{\exp(-\beta E_0) + \exp(\beta \Delta E)}.
\end{equation}
When $E_0 = 0$, detailed balance is restored with $\frac{\omega(\sigma \to -\sigma)}{\omega(-\sigma \to \sigma)} = \exp(-\beta \Delta E)$. For $E_0 = 4$, $\exp(-\beta E_0) = \exp(-4\beta)$ is a small number, especially at higher $\beta$. The dynamic is dominated by the $\Delta E$ term. Even if the system is in NESS, the active drive allows it to behave very similarly to a cold equilibrium system. Hence, $\etaInf$ differs only slightly from $\eta$. On the other hand, when $E_0 = -4$, $\exp(-\beta E_0) = \exp(4\beta)$ is a large constant that dominates the spin-flip dynamic. The spin-flip rate is pushed towards 1, meaning the system behaves as if it were subjected to infinite-temperature noise. The active drive causes large fluctuations that break the fluctuation-response relation. Hence, $\etaInf$ increases sharply while $\eta$ remains low.

\begin{figure}[h]
    \centering
    \includegraphics[width=0.48\textwidth]{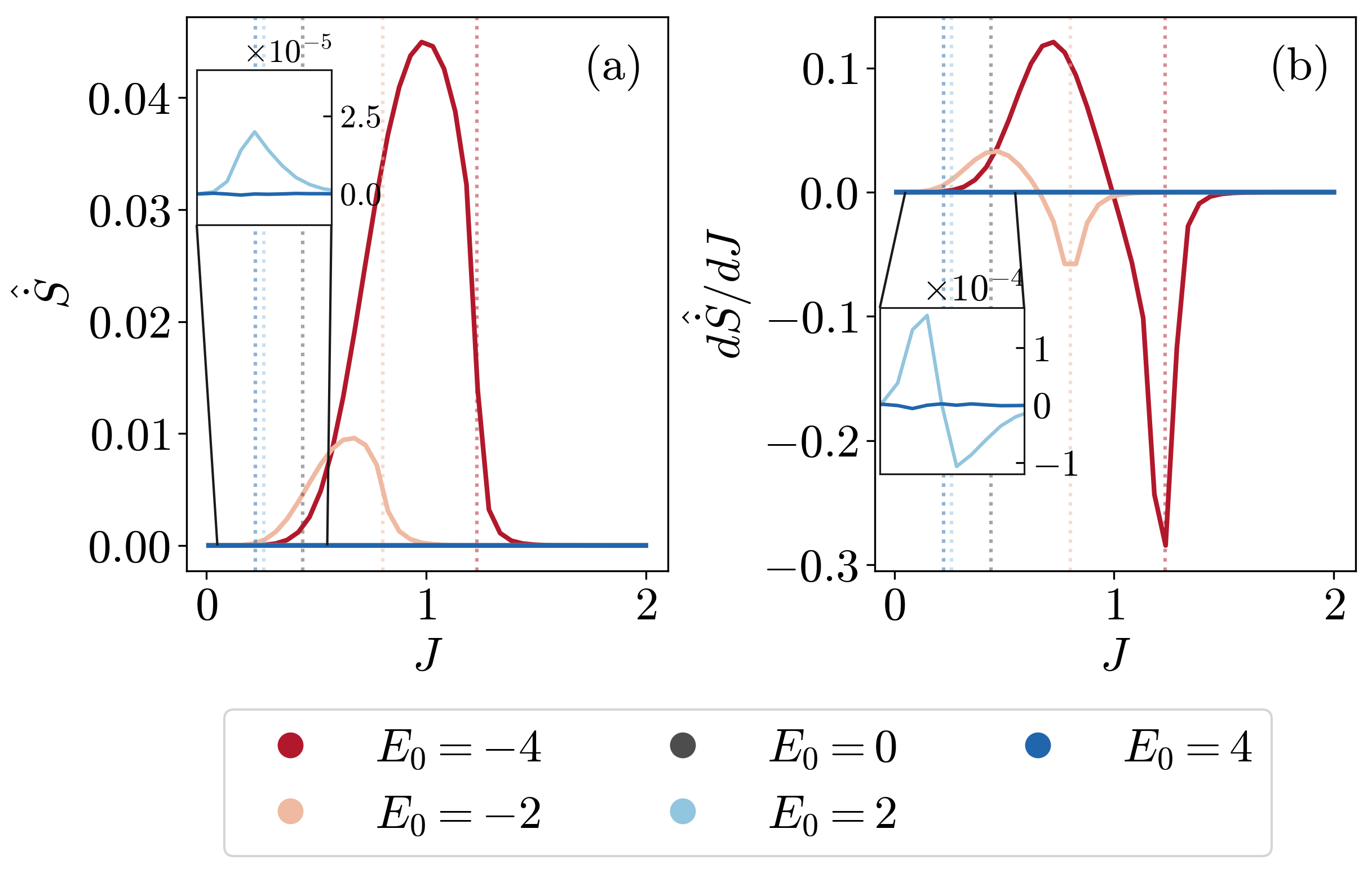}
    \caption{\textbf{Steady state entropy production rate ($\dot{S}$) and its derivative with respect to $J$}. \textbf{Left:} Entropy production rate against $J$ for different drives $E_0$, computed for steady state and normalised by the total number of particles. Inset: zoom in on $\dot{S}$ for $E_0>0$ near the transition points. \textbf{Right:} Derivative of NESS entropy production rate with respect to $J$, for different drives $E_0$. Inset: zoom in on $d\dot{S}/dJ$ for $E_0>0$ near the transition points. Lattice size: $100 \times 100$.}
    \label{fig:pim_epr_depr_v_J}
\end{figure}

The steady state entropy production rate is computed for different drives $E_0$ (Figure~\ref{fig:pim_epr_depr_v_J}(a)). Unlike thermodynamic efficiency or inferential efficiency, the entropy production rate is not maximised at the critical point. Instead, it exhibits a sharp decrease as the system transitions from the disordered phase to the ordered phase. This behaviour is clearer in its first derivative with respect to $J$, which shows a cusp at the transition point (Figure~\ref{fig:pim_epr_depr_v_J}(b)). 

\begin{figure}[h]
    \centering
    \includegraphics[width=0.48\textwidth]{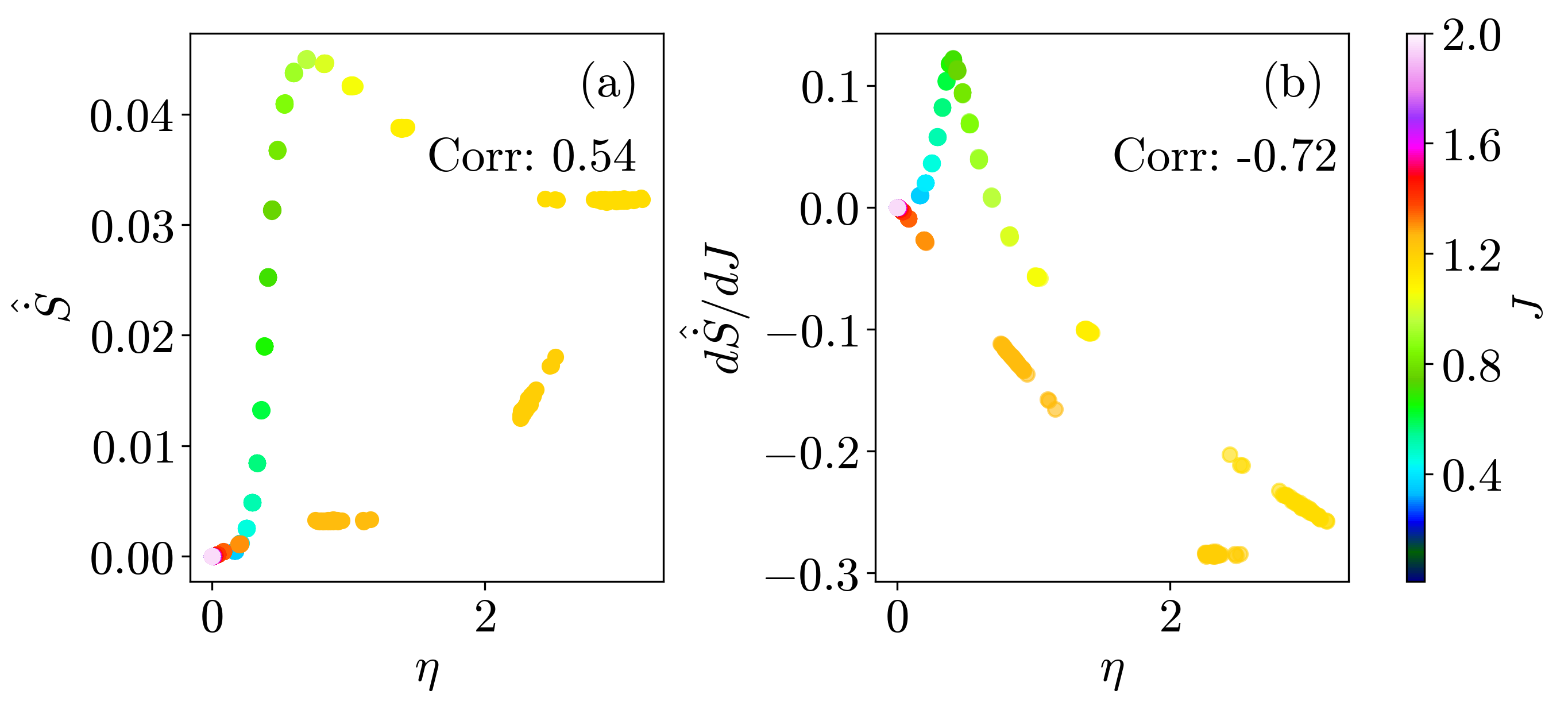}
    \caption{\textbf{Comparison of steady state entropy production rate and its derivative with thermodynamic efficiency for $E_0=-4$}. The marker colour indicates $J$. Each point represents one simulation; for each $J$, we show 48 simulations. Because the system is in steady state, many points overlap. \textbf{(a)} Entropy production rate $\dot{S}(J)$ compared with thermodynamic efficiency $\eta(J)$. \textbf{(b)} Derivative of entropy production rate $d\dot{S}(J)/dJ$ compared with thermodynamic efficiency $\eta(J)$. Lattice size: $100 \times 100$.}
    \label{fig:pim_epr_depr_v_eta}
\end{figure}

We next compare the entropy production rate with thermodynamic efficiency for $E_0=-4$ (Figure~\ref{fig:pim_epr_depr_v_eta}(a)). The scatter plot forms a loop-like trajectory rather than showing direct proportionality (correlation coefficient $\approx 0.54$). The colour map of $J$ indicates that the entropy production rate reaches maximum at $J \approx 0.98$ (green/yellow data points), before the system reaches optimal efficiency near the critical point $J_c \approx 1.23$. 

Figure~\ref{fig:pim_epr_depr_v_eta}(b) compares the derivative of entropy production rate with thermodynamic efficiency instead, and reveals a stronger (negative) correlation (correlation coefficient $\approx -0.72$). When $\eta(J)$ maximises at the transition point from disordered to ordered phase, $d\dot{S}/dJ$ reaches a negative minimum, corresponding to a sharp decrease of entropy production rate. When $\eta(J)$ is low, $d\dot{S}/dJ$ is also relatively small. The results for other values of $E_0$ are shown in Appendix \ref{ap:PIM_simulation}.

\begin{figure*}  
    \centering
    \includegraphics[width=0.95\textwidth]{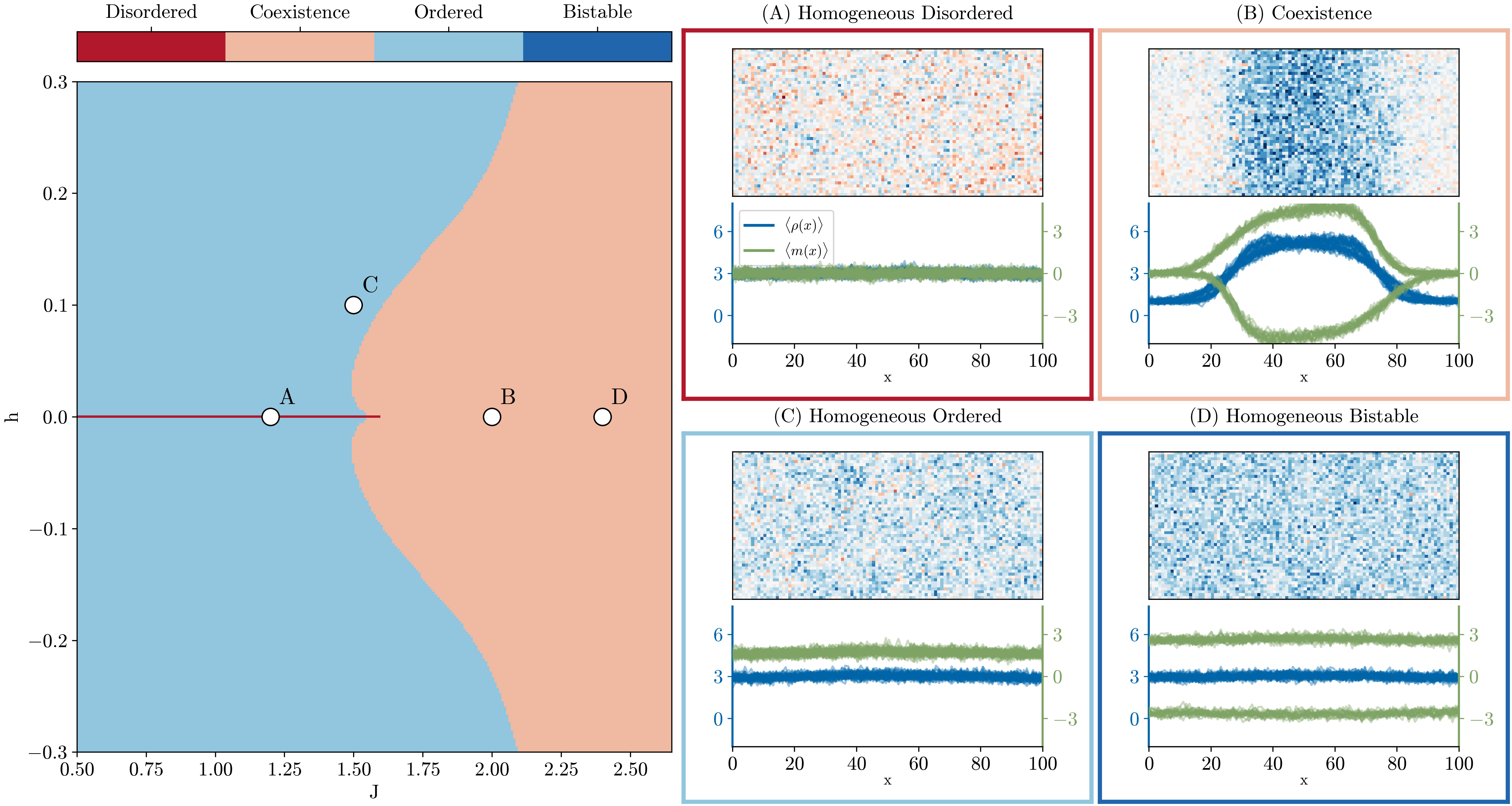}
    \caption{\textbf{Phase plot for active Ising model and example profiles}. \textbf{Left:} phase plot in the $h-J$ plane based on the refined mean-field model. \textbf{Right:} example simulation profiles corresponding to points A-D indicated in the phase plot. The homogeneous bistable phase is absent from the refined mean-field model prediction because the approximation breaks down at large $J$, and point D is incorrectly predicted as in coexistence phase by the refined mean-field model. Lattice size=$100\times 100,\rho_0=3, \beta=1, \epsilon=1$ for simulations of points A-D.}
    \label{fig:AIM_phase}
\end{figure*}

\subsection{Active Ising model}
We begin with the hydrodynamic description of the active Ising model, extended to include coupling strength $J$ and external magnetic field $h$. In a continuum limit, local density $\rho_i$ and local magnetisation $m_i$ are replaced by $\rho=\rho(x,t)$ and $m=m(x,t)$, where $x, t$ denotes the position and time, respectively. The spatio-temporal evolution of $\rho(x,t)$ and $m(x,t)$ is then governed by the hydrodynamic equations, which can be derived from the microscopic updating rule using a similar approach to that in \cite{solon2015Flocking}. The exact hydrodynamic equations are:
\begin{subequations} \label{eq:aim_exactPDF}
    \begin{align}
        \partial_t \langle \rho \rangle &= D\Delta \langle \rho \rangle - v \partial_x \langle m \rangle \\
        \partial_t \langle m \rangle &= D\Delta \langle m \rangle - v \partial_x \langle \rho \rangle \\\notag
        &+ \langle 2 \rho \sinh[\beta(h+J\frac{m}{\rho}) - 2m\cosh[\beta(h+J\frac{m}{\rho})]\rangle
    \end{align}
\end{subequations}
where $\Delta$ is the Laplace operator. $\Delta = \partial_{xx} + \partial_{yy}$ in the two-dimensional case. $v=2D\varepsilon$ is the advection speed.

We adopt the refined mean-field approximation in \cite{solon2013Revisiting, solon2015Flocking}, as the classical mean-field approach fails to capture phase-separation in the model. Under the refined mean-field approximation, $\rho$ and $m$ are treated as independent Gaussian random variables:
\begin{equation*}
    \rho \sim \mathcal{N}(\bar{\rho}, \alpha_{\rho}\bar{\rho}), \quad m \sim \mathcal{N}(\bar{m}, \alpha_{m}\bar{\rho}),
\end{equation*}
where  $\bar{\rho}=\langle \rho(x,t)\rangle$ and $\bar{m}=\langle m(x,t)\rangle$ are the mean local density and mean local magnetisation. Their variances are proportional to the mean density, with coefficients $\alpha_{\rho}, \alpha_{m}$ that depend on $\beta, v, J, h$. The resulting refined mean-field model is:
\begin{subequations} \label{eq:aim_rfmf}
\begin{align} 
    \partial_t \rho &= D\Delta \rho - v \partial_x m \\
    \partial_t m &= D\Delta m - v \partial_x \rho + C_0(\rho) + C_1(\rho) m + \frac{C_2}{\rho} m^2 - \frac{C_3}{\rho^2} m^3
\end{align}
\end{subequations}
where 
\begin{align*}
    C_0(\rho) &= 2 \rho \sinh(\beta h) + \alpha_m C_2 \\
    C_1(\rho) &= 2[(\beta J -1 )\cosh(\beta h) - \frac{\tilde{r}}{\rho}] \\
    C_2 &= \beta J (\beta J -2)\sinh(\beta h) \\
    C_3 &= \beta^2 J^2(1-\frac{\beta J}{3})\cosh(\beta h)\\
    \tilde{r} &= \frac{3\alpha_m}{2}C_3
\end{align*}
For full derivation of the hydrodynamic equations and refined mean-field approximation, see Appendix \ref{ap:AIM_derivation}.

Figure~\ref{fig:AIM_phase} shows the phase plot predicted by the refined mean-field model, together with representative examples from each of the phases. The refined mean-field model predicted three phases. In the \textit{homogeneous disordered} phase, the particles are evenly spread out on the lattice, hence cell density $\rho(x) \approx \rho_0$, while the thermal noise overtakes spin alignment, making cell magnetisation $m(x) \approx 0$ (point A in Figure~\ref{fig:AIM_phase}). In the \textit{coexistence} phase, a high-density polarised band travels on a low-density disordered background, and there are two possible polarisations (point B in Figure~\ref{fig:AIM_phase}). In the \textit{homogeneous ordered} phase, the particles are evenly distributed on the lattice, but local magnetisation is non-zero due to the influence of the non-zero external magnetic field (point C in Figure~\ref{fig:AIM_phase}). 

However, the refined mean-field model fails to predict the fourth possible phase: \textit{homogenous bistable}, which occurs when the external magnetic field $h=0$ and the coupling strength $J$ is sufficiently high. This phase is characterised by a homogeneous density profile $\rho(x) \approx \rho_0$ and two possible magnetisation profiles $m(x) \approx \pm \rho_0$ (point D in Figure~\ref{fig:AIM_phase}). The refined mean-field model relies on truncating higher-order fluctuations and ignoring $\rho$-$m$ cross-correlations. While it correctly predicts the onset of phase separation, it becomes inaccurate in the strong-coupling regime.

\begin{figure}[h]  
    \centering
    \includegraphics[width=0.4\textwidth]{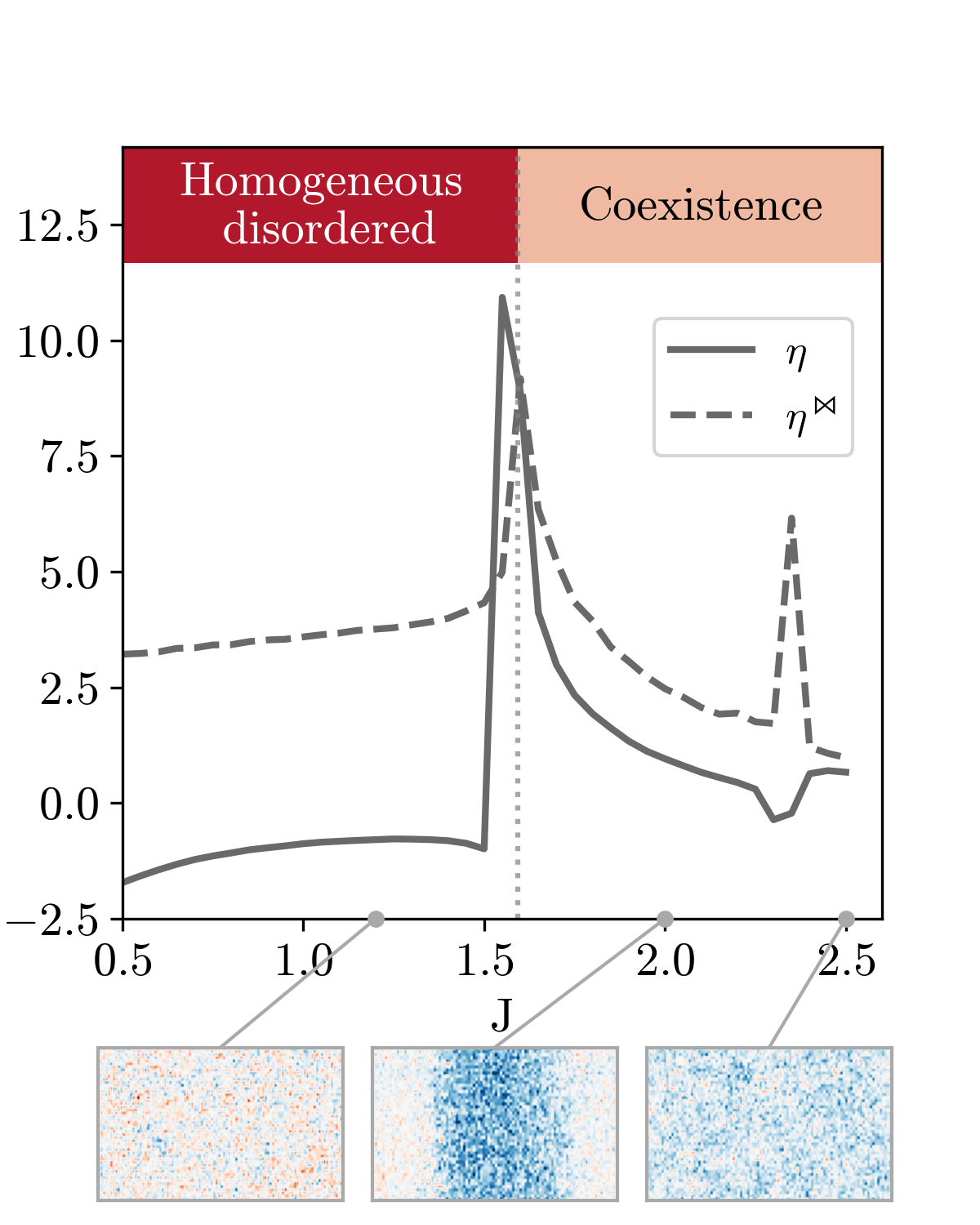}
    \caption{\textbf{Thermodynamic efficiency and inferential efficiency of active Ising model for zero external magnetic field ($h=0$).} The banner at the top shows the phases predicted by the refined mean-field model. Vertical dotted line indicates the transition point from homogeneous disordered state to coexistence state at $J_c \approx 1.5922$. Typical lattice snapshots are displayed below the x-axis for $J = 1.2$ (homogeneous disordered), $J = 2.0$ (coexistence), and $J = 2.5$ (homogeneous bistable, not identified by the refined mean-field model). Lattice size=$100\times 100, \rho_0=3, \beta=1, \epsilon=0.9$.}
    \label{fig:AIM_eta_zero_field}
\end{figure}

Figure \ref{fig:AIM_eta_zero_field} compares thermodynamic efficiency $\eta(J)$ and inferential efficiency $\etaInf(J)$ for control parameter $J$, with external magnetic field $h=0$. Both $\eta(J)$ and $\etaInf(J)$ maximise at the transition from the homogenous disordered phase to the coexistence phase. Inferential efficiency $\etaInf(J)$ is higher than thermodynamic efficiency $\eta(J)$ for most values of J, except for a narrow interval during phase transition. While active nonequilibrium driving inflates the variance of the macroscopic observable, leading to $\etaInf(J) > \eta(J)$, the singular nature of spontaneous symmetry breaking at $h=0$ tends to temporarily reverse this trend. A none-zero external field destroys the reversion and restores $\etaInf(J) > \eta(J)$ for the full range of $J$ (see Figure~\ref{fig:AIM_eta_nonzero_h} in Appendix~\ref{ap:AIM_simulation}).

Additionally, Figure \ref{fig:AIM_eta_zero_field} shows two peaks in the inferential efficiency $\etaInf(J)$, corresponding to the two phase transitions: from homogeneous disordered to coexistence, and from coexistence to homogeneous bistable. The second phase transition is not captured by the refined mean-field model, but is observed in simulations. At both transitions, inferential efficiency $\etaInf(J)$ peaks in the same direction. By contrast, thermodynamic efficiency $\eta(J)$ shows a peak at the first transition and a cusp at the second. The peak in $\eta(J)$ reflects the emergence of spatial structure in the system --- a polarised high-density band on a homogenous low-density background --- whereas the cusp reflects the sudden loss of order as the system enters the homogeneous bistable phase.

In conclusion, the result indicates that the critical regime is optimal from two different perspectives: the system is most efficient in converting consumed energy into macroscopic order, as captured by thermodynamic efficiency $\eta$, and its observables are most informative about the hidden control parameters, as measured by $\etaInf$.

\begin{figure}[h]
    \centering
    \includegraphics[width=0.33\textwidth]{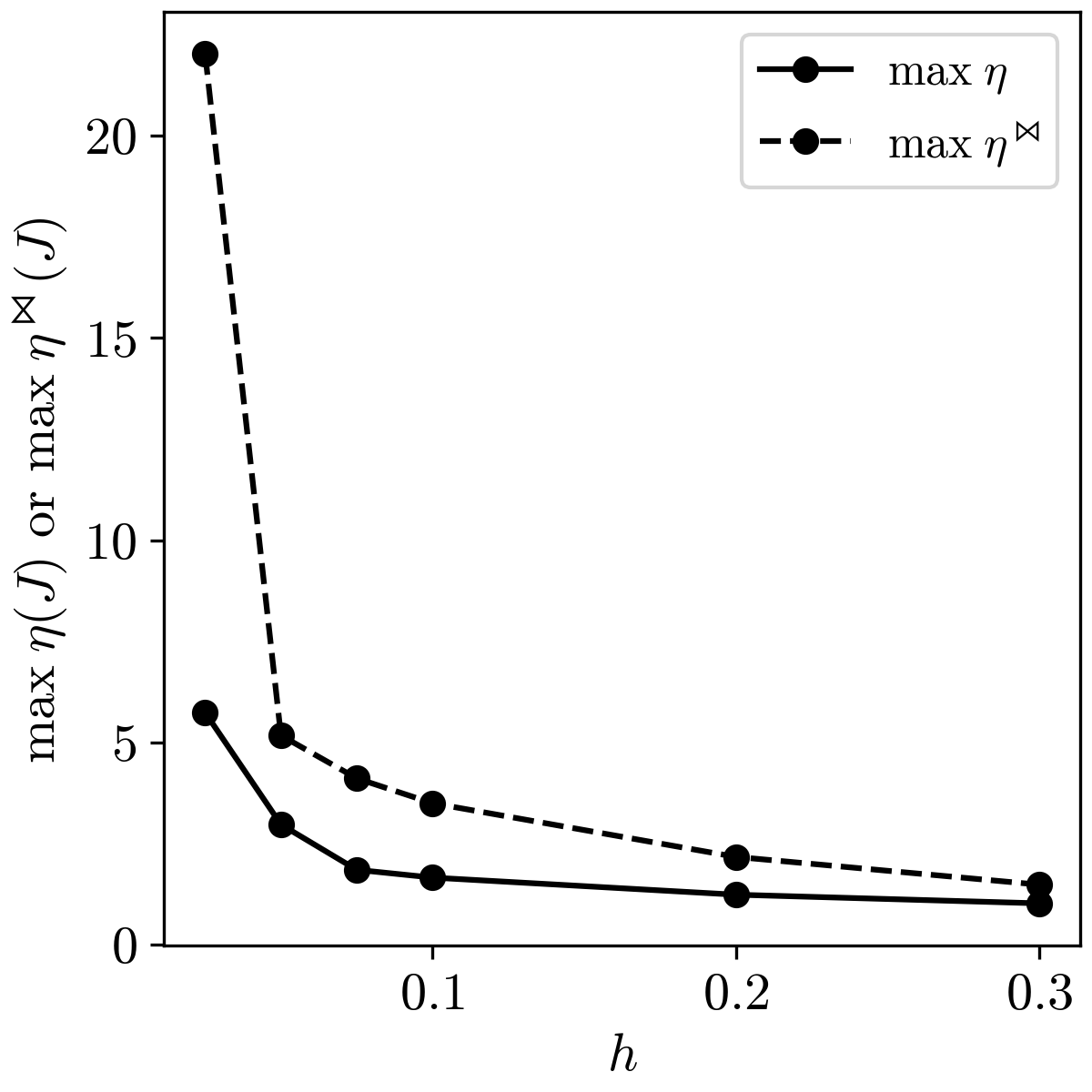}
    \caption{\textbf{Maximum inferential efficiency ($\etaInf$) and thermodynamic efficiency plotted against different external magnetic field $h$.} Lattice size=$100\times 100,\rho_0=3, \beta=1, \epsilon=0.9$.}
    \label{fig:AIM_eta_gap_h}
\end{figure}

We further compare the maximum $\eta(J)$ and $\etaInf(J)$ for external magnetic field $h > 0$ (Figure \ref{fig:AIM_eta_gap_h}). The active Ising system continuously dissipates energy via particle self-propulsion and diffusion, thereby increasing the variance of macroscopic observables. As a result, the inferential efficiency $\etaInf(J)$, which is proportional to the fluctuation of the macroscopic observable, is expected to be higher than the thermodynamic efficiency $\eta(J)$, which captures the true response of the system to the perturbation of the control parameter $J$. 

However, as the external magnetic field increases, the field term begins to dominate the dynamics, forcing all spins to align in the same direction. The system is pushed into a highly ordered, homogeneous state, and the energy penalty for any spin to flip against the field becomes high. This external constraint effectively freezes the system, suppressing both thermal and self-propulsion-driven fluctuations. The resulting decrease in the variance of the macroscopic observable weakens the violation of the fluctuation-response relation and narrows the gap between $\etaInf(J)$ and $\eta(J)$.

\begin{figure}[h]  
    \centering
    \includegraphics[width=0.4\textwidth]{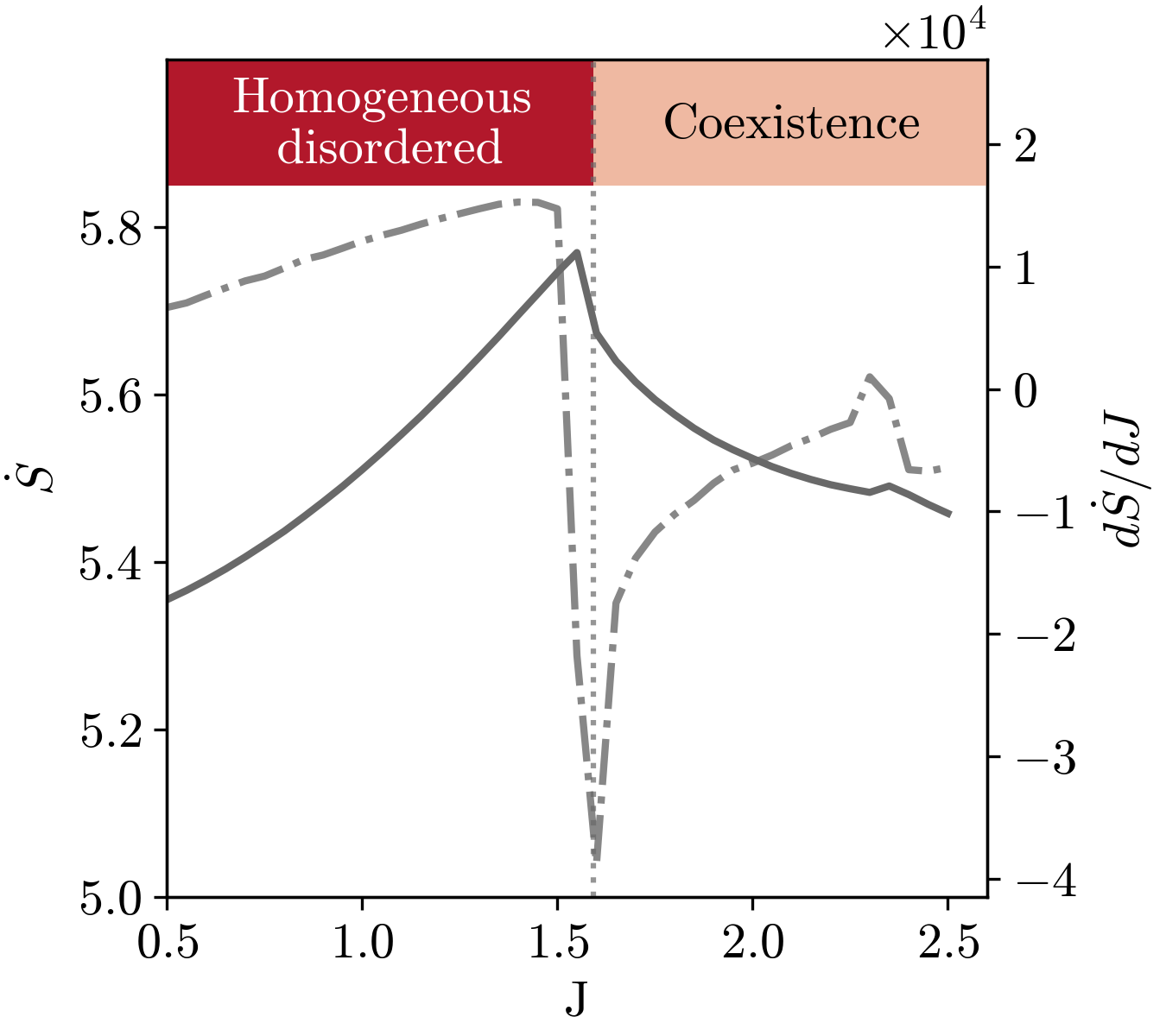}
    \caption{\textbf{Entropy production rate and its derivative for active Ising model for zero external magnetic field ($h=0$).} The banner shows the phases predicted by the refined mean-field model. Vertical dotted line indicates the transition point from homogeneous disordered state to coexistence state at $J_c \approx 1.5922$. Entropy production rate (solid curve) is normalised by the total number of particles. Its derivative with respect to $J$ is shown in dash-dot curve. Lattice size=$100\times 100, \rho_0=3, \beta=1, \epsilon=0.9$.}
    \label{fig:AIM_epr_depr_v_J}
\end{figure}

Figure~\ref{fig:AIM_epr_depr_v_J} shows the entropy production rate, $\dot{S}$, and its derivative with respect to the coupling strength, $d\dot{S}/dJ$, as functions of $J$ at $h=0$. As $J$ increases, the total entropy production rate (solid curve, primary y-axis) exhibits a bimodal curve as the system passes through three distinct phases: the homogeneous disordered phase, the coexistence phase, and the homogeneous bistable phase. The homogeneous bistable phase emerges at high $J$ when $h$ is sufficiently small, but is absent from the refined mean-field predictions. The entropy production rate $\dot{S}$ rises to a maximum at the onset of phase coexistence ($J \approx 1.5922$), then decreases before forming a much subtler second peak near the transition into the homogeneous bistable regime ($J \approx 2.3$). 

The behaviour of the derivative $d\dot{S}/dJ$ (Figure~\ref{fig:AIM_epr_depr_v_J}, dash-dot curve, secondary y-axis) reveals the two transitions more clearly. At both transition points, $d\dot{S}/dJ$ exhibits discontinuities. In contrast to thermodynamic efficiency $\eta(J)$ (Figure~\ref{fig:AIM_eta_zero_field}), which peaks at the first transition, the derivative $d\dot{S}/dJ$ is negative and forms a cusp as the system transitions from homogeneous disordered to coexistence phase. The cusp reflects a sharp decrease in entropy production rate. At the second transition, from coexistence to homogeneous bistable phase, $d\dot{S}/dJ$ shows a positive peak as the spatial structure breaks down and the system reorganises into a homogeneous state.

\section{Conclusions and discussion \label{sec:discussion}}

In this paper, we investigated the efficiency of nonequilibrium self-organising systems. We studied two collective systems that are driven out of equilibrium under different mechanisms: the persistent Ising model \cite{kumar2020Nonequilibrium}, where the constant bias $E_0$ introduced to the transition probability breaks the detailed balance condition, and the active Ising model \cite{solon2013Revisiting, solon2015Flocking}, where nonequilibrium arises from particle self-propulsion.

The key finding from this analysis is that the thermodynamic efficiency of a NESS system is maximised at the critical regime, although this regime generally differs from the equilibrium critical point. The corresponding critical control parameter values depend on the external drive. This finding extends previous results for equilibrium systems \cite{crosato2018Critical, crosato2018Thermodynamics, harding2018Thermodynamic, nigmatullin2021Thermodynamic, chen2025Why} to systems out of equilibrium. More broadly, this indicates that a collective system coordinates most efficiently in the critical regime, whether in equilibrium or NESS.

Furthermore, the divergence between thermodynamic efficiency $\eta$ and inferential efficiency $\eta^{\bowtie}$ provides a phenomenological signature of broken detailed balance and violation of fluctuation-dissipation theorem \cite{kubo1957StatisticalMechanical}. Thermodynamic efficiency $\eta$ is computed from the entropy reduction and work associated with a small perturbation of the control parameter. By contrast, inferential efficiency $\eta^{\bowtie}$ replaces this response to perturbation with equal-time covariances of the macroscopic observables measured in the unperturbed steady state. At equilibrium, fluctuation-dissipation theorem links the linear response of the system to an external perturbation ($\partial \langle X_i \rangle / \partial {\lambda_j}$) to the fluctuation of macroscopic observables about their means ($\operatorname{Cov}(X_i, X_j)$) \cite{jaynes1963Information, kivelson2024Statisticalch7}. Thus, $\eta=\eta^{\bowtie}$ follows from the equilibrium relation between spontaneous fluctuations and linear response.

Away from equilibrium, the fluctuation-response identity may not hold \cite{cugliandoloEnergy, harada2004Fluctuationresponse, harada2005Equality}. Our results show that the gap between $\eta$ and $\eta^{\bowtie}$ increases with stronger active driving, suggesting that the nonequilibrium driving disrupts the equilibrium link between fluctuation and linear response. This interpretation is consistent with the Harada-Sasa relation, which connects dynamical fluctuation-response violation to energy dissipation in Langevin systems \cite{harada2005Equality}. However, our gap is constructed from static susceptibilities and equal-time covariances, rather than dynamical correlation and response functions in the frequency domain. It should therefore be interpreted not as a quantitative measure of distance from equilibrium, but as a phenomenological signature of nonequilibrium driving.

Our finding that thermodynamic efficiency maximises at nonequilibrium phase transitions provides a new perspective when contrasted with the behaviour of NESS entropy production rate. Recent studies show that nonequilibrium phase transitions can be characterised by the entropy production rate and its first derivative \cite{tome2012Entropy, nguyen2018Phase, noa2019Entropy}. We evaluated these two quantities here and compared them with thermodynamic efficiency. Thermodynamic efficiency shows a moderate correlation with the entropy production rate, but follows more closely with the derivative of entropy production rate: high thermodynamic efficiency coincides with large $|d\dot{S}(J)/dJ|$. During the phase transition from disordered to ordered, thermodynamic efficiency is maximised and positive, reflecting the formation of internal structure with little work input. At the same transition, derivative of entropy production rate reaches a negative minimum, indicating a sharp reduction in the entropy production rate. These results complement the previous studies by framing the critical regime not only as a singularity in dissipative fluxes, but as a regime of optimal work-to-order conversion for self-organisation.

Our findings both confirm and extend the findings of the earlier lattice model studies \cite{solon2013Revisiting, solon2015Flocking, kumar2020Nonequilibrium}. For the persistent Ising model, Kumar and Dasgupta used temperature as the control parameter and showed that the specific heat per spin diverges from the derivative $d\langle E \rangle / dT$ when $E_0 \neq 0$. Using the coupling strength $J$ as the control parameter, we have a similar observation and further reveal a thermodynamic asymmetry: when $E_0<0$ (self-excited spin-flip dynamics), the system's total fluctuation is higher than the rate at which its configurational entropy reduces. This suggests that a fraction of the fluctuation induced by the parameter change does not translate into stable structural order and is associated with the dissipated heat required to keep the system out of equilibrium. 

For the active Ising model, Solon and Tailleur \cite{solon2013Revisiting, solon2015Flocking} demonstrated that a liquid-gas phase separation occurs by tuning average density $\rho_0$ and temperature $T$ to the appropriate regime. In this study, we used the coupling strength $J$ and the external magnetic field $h$ as control parameters and recovered the coexistence phase while uncovering an additional homogeneous ordered phase induced by a non-zero $h$. Crucially, we showed that a strong external field dampens the nonequilibrium effect from active self-propulsion, shrinking the violation of the fluctuation-response relation and forcing the system back to a pseudo-equilibrium state.

This study is subject to several limitations that point toward directions for future work. First, the external drives in both models are constant in time. While this formulation allows the analysis to focus on NESS systems, real collective systems are more often exposed to time-dependent or fluctuating drives. Extending the analysis to explicitly account for a time-varying drive (e.g. see \cite{tome1990Dynamic, chakrabarti1999Dynamic, seara2021Irreversibility, quintana2023Experimental}) would allow one to investigate transient responses, hysteresis, and path-dependent effects that cannot be captured within the current framework. 

Second, the model assumes homogeneous agents with identical coupling strengths and identical responses to the external drive. In many natural and social systems, agents are heterogeneous and may pursue competing or partially aligned objectives \cite{kim2010Quantum, avni2025Nonreciprocal, mangeat2025Emergent}. Introducing heterogeneity in coupling strengths, local interaction rules, or agent-specific drive parameters would enable the study of how diversity and competing influences shape collective behaviour under nonequilibrium conditions. However, applying our thermodynamic efficiency measure to systems with nonreciprocal or competing interactions requires adapting how generalised work is measured. Because such systems lack a global Hamiltonian, work cannot be evaluated from the Hamiltonian as in this study. Instead, generalised work must be computed directly from the state-dependent energy changes induced by the time-dependent protocol \cite[~ch.3]{shiraishi2023Introduction}.

Finally, our analysis focuses on steady-state behaviour after the system has fully relaxed. This excludes regimes in which the dynamics are fast relative to the system's relaxation timescale. Investigating such regimes would be relevant for understanding collective behaviour under rapid environmental change, where steady-state assumptions may no longer hold. A key challenge is that estimating entropy and work rates during transients is more difficult, limiting the applicability of current thermodynamic efficiency measures.

In conclusion, this study showed that in a driven, nonequilibrium setting, a self-organising collective system achieves optimal thermodynamic efficiency at the critical regime. Unlike classical heat engines, which convert energy into mechanical work, biological and active systems operate as engines of order generation: they dissipate energy to maintain low-entropy, highly structured states (e.g., flocking or tissue formation). The information-theoretic thermodynamic efficiency $\eta$ measures this 'work-to-order' conversion. Our observation that $\eta$ maximises at the nonequilibrium phase transition suggests that active systems translate work into macroscopic functional structure most efficiently near criticality. From this perspective, the ubiquity of critical behaviour in self-organising systems may be understood through a broader \textit{Principle of Super-efficiency} \cite{chen2025Why}: collective systems tend to self-organise toward regimes that maximise the conversion of energetic expenditure into coordinated behaviour, and these regimes coincide with the critical regime. This study advances the understanding of the efficiency of self-organising systems out of equilibrium and the prevalence of criticality in nature.

\section*{Acknowledgement}
The authors would like to thank David Wolpert for the insightful discussion on stochastic thermodynamics. Q.C. would like to thank Emanuele Crosato for the comments on computing generalised work for equilibrium systems, Ibrahim Al-Azki for the discussions on entropy estimation algorithms, and Carlos Gershenson for the idea of thermodynamic efficiency in heterogeneous agents. Q.C. is supported by the University of Sydney Postgraduate Award (UPA). We wish to acknowledge the support of the Sydney Informatics Hub at the University of Sydney and the National Computing Infrastructure (NCI) Australia for high-performance computing resources that have contributed to the research results reported within this paper.

\appendix


\section{Simulation of persistent Ising model} \label{ap:PIM_simulation}

We simulate the persistent Ising model on a $100 \times 100$ lattice with periodic boundary conditions for the main analysis. In each Monte Carlo step, a particle is selected and its spin is updated using the modified Glauber rule: 
\begin{equation*}
    \omega(\sigma \to -\sigma) = \frac{1}{1+\exp(\beta \Delta\tilde{E}_{\sigma \rightarrow -\sigma})},
\end{equation*}
as specified in Eqs.~\eqref{eq:pim_E_tilde} and ~\eqref{eq:pim_glauber}. One sweep consists of $N = 100 \times 100$ Monte Carlo steps. The system relaxes to a steady state after approximately $80,000$ sweeps. We take one sample each sweep. For each value of J, we run $48$ independent simulations. After the system is fully relaxed, $5000$ samples are taken for each simulation to compute entropy and average work. The final values are then obtained by averaging over all simulations. The system is initialised at a fully ordered state to avoid being trapped at metastable states at high coupling strength.

Entropy is computed using the Kikuchi approximation \cite{kikuchi1951Theory}:
\begin{equation*}
    S = S_4 - 2 S_2 + S1,
\end{equation*}
where $S_k$ is the entropy of the size-k sublattices. 

The work associated with the small perturbation $\delta J$ of the control parameter is computed using the change in Hamiltonian \cite[Sec 3.2.2]{seifert2025Stochastic} before and after the perturbation, assuming that the system's configuration $\underline{\sigma}$ does not change during the perturbation:
\begin{align*}
    W_{J \to J+\delta J} &= \mathcal{H}(\underline{\sigma}; J+\delta J) - \mathcal{H}(\underline{\sigma}; J) \\
    &= -(J+\delta J) \sum_{\langle ij \rangle}\sigma_i\sigma_j + J \sum_{\langle ij \rangle}\sigma_i\sigma_j \\
    &= -\delta J \sum_{\langle ij \rangle}\sigma_i\sigma_j
\end{align*}

\begin{figure}[H]
     \centering
     \begin{subfigure}[b]{0.5\textwidth}
         \centering
         \includegraphics[width=\textwidth]{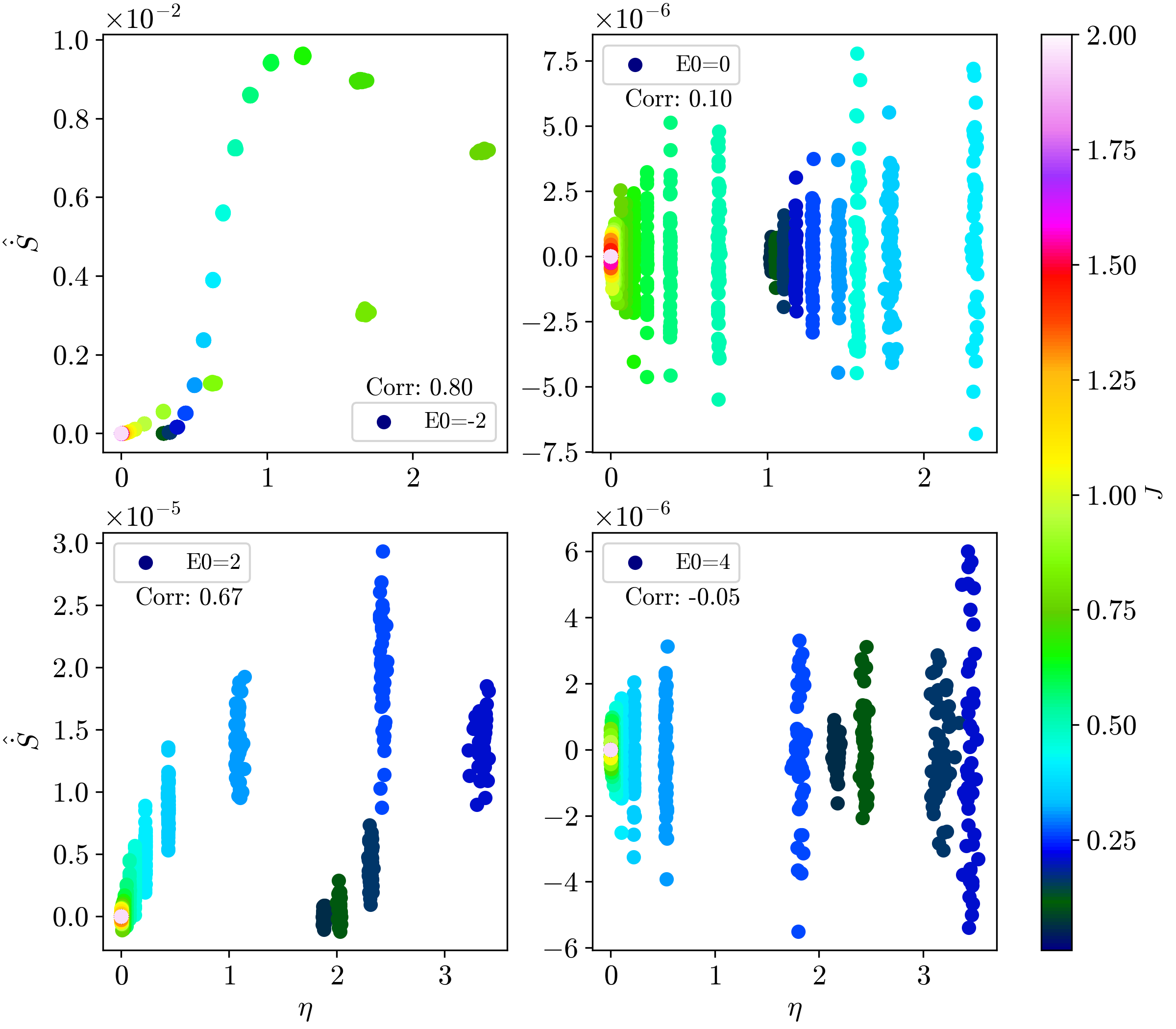}
         \caption{$\eta$ vs $\hat{\dot S}$}
     \end{subfigure}
     \vfill
     \begin{subfigure}[b]{0.5\textwidth}
         \centering
         \includegraphics[width=\textwidth]{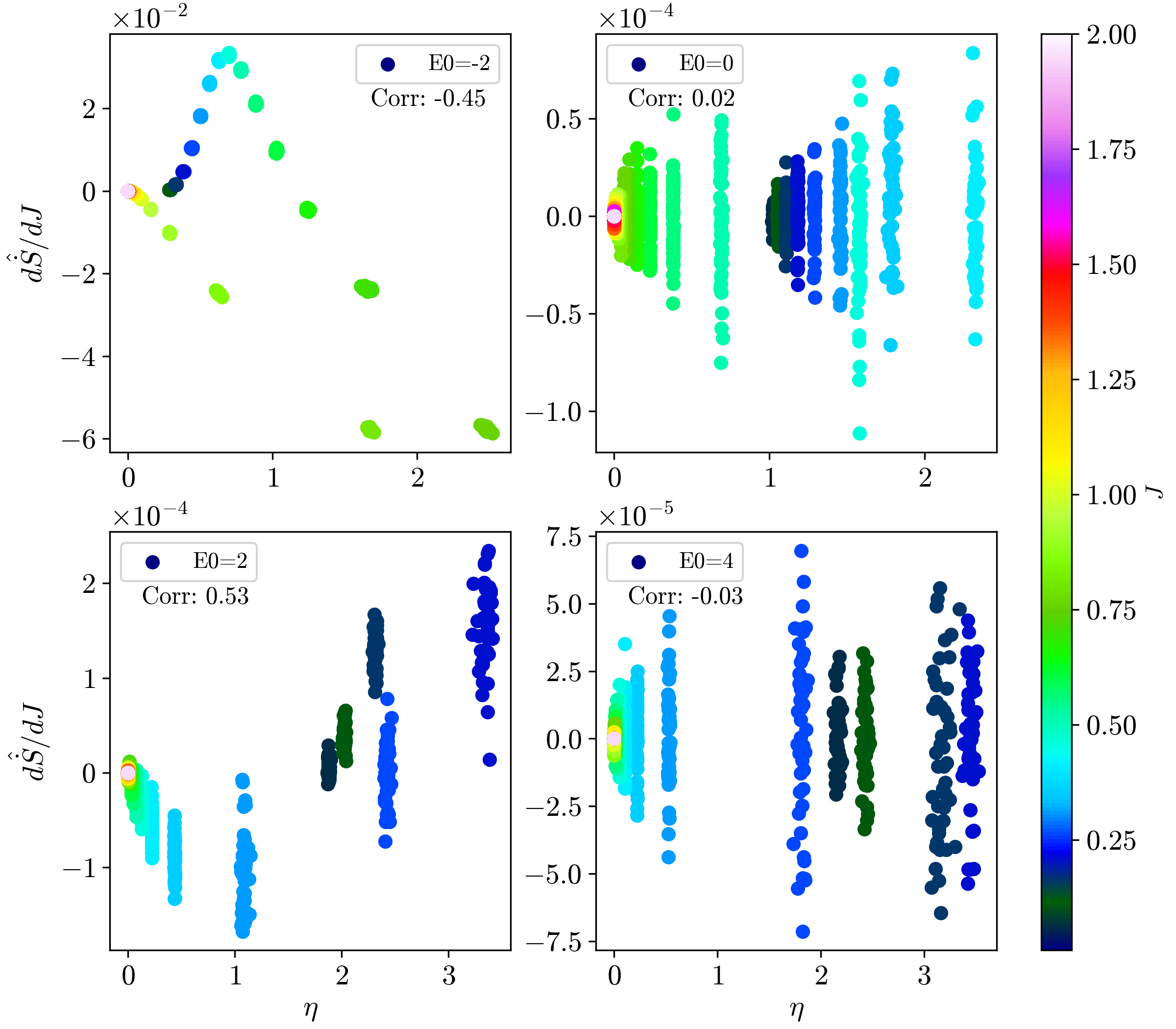}
         \caption{$\eta$ vs $d\hat{\dot S}/dJ$}
     \end{subfigure}
     \caption{\textbf{Comparisons of thermodynamic efficiency with entropy production rate and its first derivative for different values of $E_0$}. The marker colour indicates $J$. Each point represents one simulation; for each $J$, we show 48 simulations. Because the system is in steady state, many points overlap. \textbf{(a)} Entropy production rate $\dot{S}(J)$ compared with thermodynamic efficiency $\eta(J)$. \textbf{(b)} Derivative of entropy production rate $d\dot{S}(J)/dJ$ compared with thermodynamic efficiency $\eta(J)$. Lattice size: $100 \times 100$.}
     \label{fig:PIM_epr_depr_v_eta_combined}
\end{figure}

We estimate the steady-state entropy production rate from the log ratio of forward and reverse single-spin-flip transition rates along sampled trajectories after the system has fully relaxed. The entropy production rate is computed as:
\begin{equation*}
    \hat{\dot S} = \frac{1}{\tau_{\mathrm{obs}}}\sum_{n}\ln\frac{W(C_n\to C_{n+1})}{W(C_{n+1}\to C_n)},
\end{equation*}
where $C_n$ represents the configuration of the focal spin and its nearest four neighbours at step $n$, and $\tau_{\mathrm{obs}}$ is measured in Monte Carlo steps. A rejected update is treated as a transition to the same configuration, hence contributes zero to the sum.

Figure~\ref{fig:PIM_epr_depr_v_eta_combined} compares thermodynamic efficiency $\eta$ with entropy production rate, ${\dot S}$, and its first derivative, $d{\dot S}/dJ$. The results for $E_0=-2$ are consistent with those in Figure~\ref{fig:pim_epr_depr_v_eta}, where the system is driven at $E_0=-4$. Where $E_0=0$, entropy production rate fluctuates around zero, and we observe no correlation between ${\dot S}$ and $\eta$. When $E_0>0$, nonequilibrium drive suppresses spin-flip activities, leading to very low entropy production. Thus, it is unclear whether the exhibited structures reflect a genuine correlation or simulation noise.

\section{Derivation of refined mean-field equations for modified active Ising model}
\label{ap:AIM_derivation}

In this section, we derive the refined mean-field equations for the active Ising model with coupling strength $J$ and external magnetic field $h$ as described in Section~\ref{sec:model_AIM}. The time evolution of local density $\rho = \rho(x,t)$ and local magnetisation $m = m(x,t)$ is described by Eq.~\eqref{eq:aim_exactPDF}, which we repeat here:
\begin{subequations}
    \begin{align*}
    \partial_t \rhoMean &= D\Delta \rhoMean - v \partial_x \mMean \\
    \partial_t \mMean &= D\Delta \mMean - v \partial_x \rhoMean \\
    &\quad + \langle 2 \rho \sinh{[\beta(h + J\frac{m}{\rho})]} - 2 m \cosh{[\beta(h + J\frac{m}{\rho})]}\rangle
\end{align*}
\end{subequations}
where $D$ is the diffusion coefficient and $v = 2D\varepsilon$ is the advection speed associated with the self-propulsion dynamics.

To obtain the refined mean-field model, we assume that:
\begin{itemize}
    \item $\rho$ and $m$ are statistically independent, so that $\langle f(m)g(\rho)\rangle \approx \langle f(m) \rangle\langle g(\rho) \rangle$ for any function $f(.), g(.)$.
    \item $\rho$ and $m$ follow Gaussian distributions with means $\bar{\rho}$ and $\bar{m}$, respectively, and variances proportional to the mean density: $\rho \sim \mathcal{N}(\bar{\rho}, \alpha_{\rho}\bar{\rho}), \quad m \sim \mathcal{N}(\bar{m}, \alpha_{m}\bar{\rho})$. $\alpha_m$ and $\alpha_\rho$ are coefficients calibrated from the simulations.
\end{itemize} 

We use the following shorthand for the derivation:
\begin{align*}
    H &\equiv \beta h \\
    u(m, \rho) & \equiv \beta J\frac{m}{\rho} \\
    I & \equiv 2 \rho \sinh{(H+u)} - 2 m \cosh{(H+u)}
\end{align*}

Note that the non-linear term can be factorised into:
\begin{align} \label{eq:aim_interaction0}
    I =  &\;\sinh(H)\left[ 2\rho \cosh(u) - 2m\sinh(u)\right] \\ \notag
    + &\;\cosh(H)\left[ 2\rho \sinh(u) - 2m\cosh(u)\right]
\end{align}

Assuming $u=\beta J\frac{m}{\rho}$ is small enough, we can replace hyperbolic functions with polynomial expansions:
\begin{align*}
    \sinh(x) &= x + \frac{x^3}{6} + O(x^5) \\
    \cosh(x) &= 1 + \frac{x^2}{2} + O(x^4).
\end{align*}
Substituting the polynomial expansion into Eq.~\eqref{eq:aim_interaction0} and collecting like terms in powers of $m$ yields:
\begin{align}
    I = \;& 2\rho \sinh(H) + (2\beta J - 2)\cosh(H)\,m \\ \notag
    + \;&\frac{\beta J}{\rho}(\beta J -2)\sinh(H)\,m^2 \\ \notag
    - &\left( \frac{\beta J}{\rho}\right)^2\left(1-\frac{\beta J}{3}\right)\cosh(H)\,m^3 + O(m^4)\\ \notag
\end{align}
Taking ensemble average of $I$ while assuming no interaction between $\rho$ and $m$ yields:
\begin{align} \label{eq:aim_interaction1}
    \langle I \rangle \approx &\; 2\rhoMean \sinh(H) + (2\beta J - 2)\cosh(H)\mMean \\ \notag
    + &\;C_2\langle \frac{1}{\rho}\rangle\langle m^2 \rangle 
    - C_3\left\langle\frac{1}{\rho^2}\right\rangle\langle m^3 \rangle
\end{align}
where 
\begin{align}
    C_2 &= \beta J(\beta J -2)\sinh(H) \\
    C_3 &= (\beta J)^2\left(1-\frac{\beta J}{3}\right)\cosh(H)
\end{align}

Gaussian moments of $m$ are:
\begin{align*}
    \mMean &= \bar{m} \\
    \langle m^2 \rangle &= \bar{m}^2 + \alpha_m\bar{\rho} \\
    \langle m^3 \rangle &= \bar{m}^3 + 3 \alpha_m \bar{\rho} \bar{m}
\end{align*}
To obtain moments of $\frac{1}{\rho}$, we assume that $\rho = \bar{\rho} + \delta \rho$ where$\frac{\delta \rho}{\bar{\rho}}$ is a small. For the first inverse moment:
\begin{align*}
    \frac{1}{\rho} &=\frac{1}{\bar{\rho}} \left( 1 + \frac{\delta \rho}{\bar{\rho}}\right)^{-1} \\
    &=\frac{1}{\bar{\rho}} - \frac{\delta\rho}{\bar{\rho}^2} + \frac{\delta \rho ^2}{\bar{\rho}^3} +O(\bar{\rho}^{-4})
\end{align*}
Taking average and dropping higher order terms $O(\bar{\rho}^{-4})$ yields:
\begin{equation}
    \left\langle\frac{1}{\rho}\right\rangle \approx\frac{1}{\bar{\rho}} + \frac{\alpha_\rho}{\bar{\rho}^2}
\end{equation}

The second inverse moment is given by:
\begin{align*}
    \frac{1}{\rho} &=\frac{1}{\bar{\rho}^2} \left( 1 + \frac{\delta \rho}{\bar{\rho}}\right)^{-2} \\
    &=\frac{1}{\bar{\rho}^2} - 2\frac{\delta\rho}{\bar{\rho}^3} + 3\frac{\delta \rho ^2}{\bar{\rho}^4} +O(\bar{\rho}^{-5})
\end{align*}
Taking average and dropping higher order terms $O(\bar{\rho}^{-5})$ yields:
\begin{equation}
    \left\langle\frac{1}{\rho^2}\right\rangle \approx\frac{1}{\bar{\rho}^2} + \frac{3\alpha_\rho}{\bar{\rho}^3}
\end{equation}

Substituting Gaussian moments of $m$ and $\rho$ into the non-linear term $\langle I \rangle$ in Eq.~\eqref{eq:aim_interaction1} and keeping only leading terms yields:
\begin{equation} \label{eq:aim_interaction2}
    \langle I \rangle \approx C_0(\bar{\rho}) + C_1(\bar{\rho}) \bar{m} + \frac{C_2}{\bar{\rho}} \bar{m}^2 - \frac{C_3}{\bar{\rho}^2} \bar{m}^3,
\end{equation}
where 
\begin{align*}
    C_0(\bar{\rho}) &= 2 \bar{\rho} \sinh(\beta h) + \alpha_m C_2 \\
    C_1(\bar{\rho}) &= 2[(\beta J -1 )\cosh(\beta h) - \frac{\tilde{r}}{\bar{\rho}}] \\
    C_2 &= \beta J (\beta J -2)\sinh(\beta h) \\
    C_3 &= \beta^2 J^2(1-\frac{\beta J}{3})\cosh(\beta h)\\
    \tilde{r} &= \frac{3\alpha_m}{2}C_3
\end{align*}
Substituting the simplified non-linear term into the exact hydrodynamic equations, Eq.~\eqref{eq:aim_exactPDF}, gives the refined mean-field equations, Eq.~\eqref{eq:aim_rfmf}. For simplicity, we replaced $\bar{\rho}$ and $\bar{m}$ with $\rho$ and $m$ in the final mean-field model.

\begin{figure}[h]  
    \centering
    \includegraphics[width=0.4\textwidth]{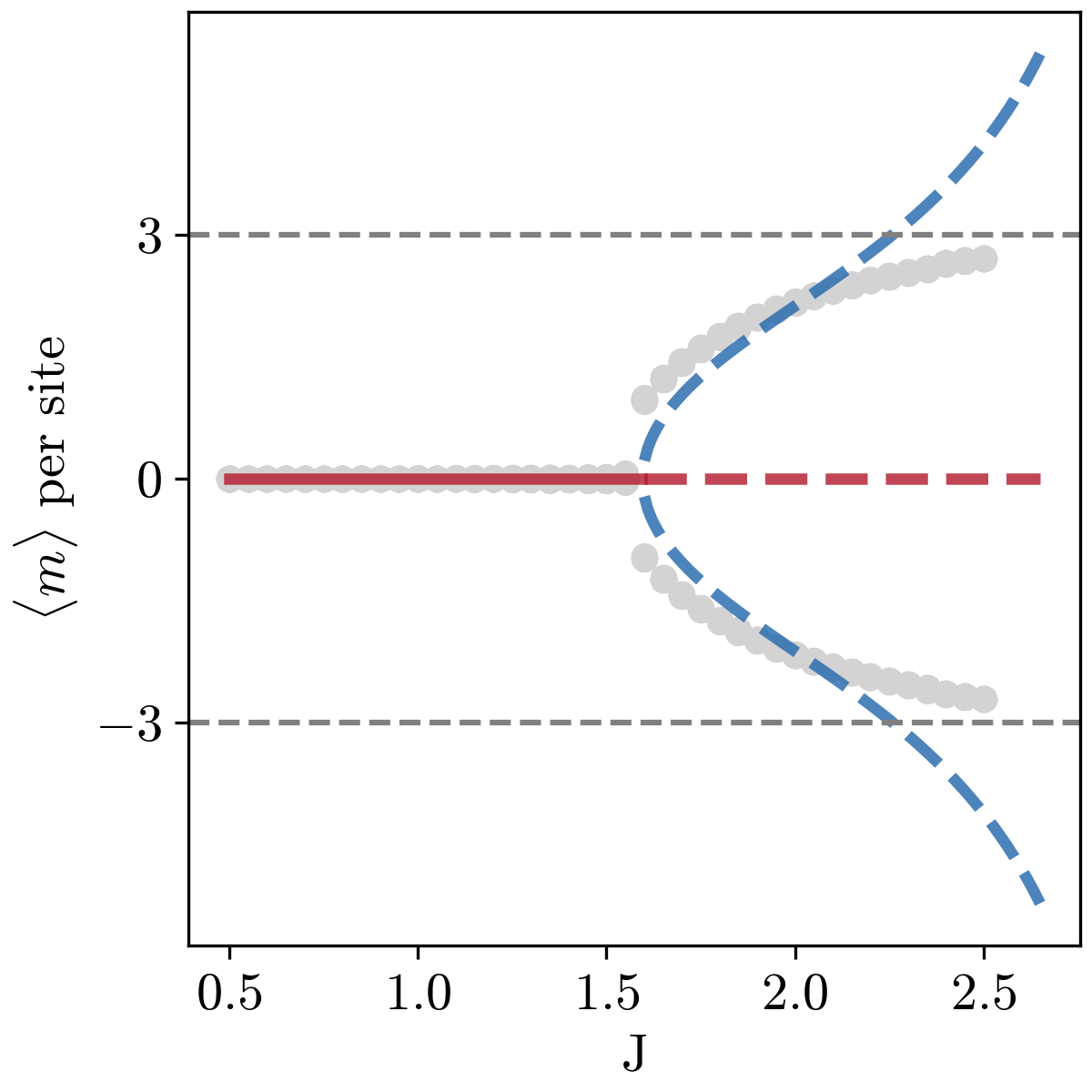}
    \caption{\textbf{Homogeneous solutions to refined mean-field model and simulation results for zero external magnetic field ($h=0$).} Solid and dashed lines denote stable and unstable homogeneous solutions, respectively. The disordered branch is indicated in red, and ordered branches are in blue. Grey dots represent 48 simulation runs for each value of $J$. For simulations, lattice size=$100\times 100, \rho_0=3, \beta=1, \epsilon=1.0$.}
    \label{fig:AIM_bifurcation}
\end{figure}

Figure \ref{fig:AIM_bifurcation} shows the homogenous solutions $(\rho_0, m_0)$ for the refined mean-field model compared against simulation results for $h=0$, using hyperparameters $\rho_0=3, \beta=1, \epsilon=1.0$. The solutions are computed by numerically solving the cubic equation:
\begin{equation}
    0 = C_0(\rho_0) + C_1(\rho_0) m + \frac{C_2}{\rho_0} m^2 - \frac{C_3}{\rho_0^2} m^3.
\end{equation}
In this case, the bifurcation happens at $J_c \approx 1.57$. For $J<1.57$, there is one stable homogeneous solution $m_0 = 0$, which is the disordered solution (solid red line). For $J>1.57$, there are three homogeneous solutions: one disordered and two ordered. All three solutions are predicted as unstable (dashed lines). Note that the ordered branches also diverge as $J>2$, while in reality they should saturate at $\pm\rho_0$, since for a homogeneous and ordered system, each cell on average has $\rho_0$ particles and the maximum magnetisation is $\pm \rho_0$ for an average cell. The divergence of the branches further indicates that the refined mean-field approximation becomes inaccurate at large $J$. 

Importantly, the scatter plot of mean magnetisation $\langle m \rangle$ computed from simulations cannot distinguish the coexistence phase and the homogenous bistable phase. In the bifurcation plot, the mean magnetisation (grey dots) is computed by summing the spins of all particles and dividing by the total number of particles. This scalar quantity masks the spatial structure presented in the coexistence phase. For $J \in [1.57, 2.3]$, the system exhibits a high-density polarised band travelling on a low-density disordered background. The mean magnetisation $\langle m \rangle$ is therefore a weighted average of the two regions. The refined mean-field model correctly identifies this regime by predicting that the homogeneous disordered solution is unstable (blue dashed line). However, for $J>2.3$, the system enters the homogeneous bistable phase, which the refined mean-field model fails to predict unless higher order terms are retained in the approximation.

\section{Stability analysis of active Ising model}\label{ap:AIM_stability}
In this section, we perform stability analysis for the active Ising model with coupling strength $J$ and external magnetic field $h$ as described in Section~\ref{sec:model_AIM}. 

Let the homogeneous steady state solutions be $(\rho_0, m_0)$. These solutions satisfy the following conditions:
\begin{align*}
    \partial_t \rho|_{\rho=\rho_0} &= 0, \quad \partial_x \rho|_{\rho=\rho_0}=0, \quad \Delta \rho|_{\rho=\rho_0}=0 \\
    \partial_t m|_{m=m_0} &= 0, \quad \partial_x m|_{m=m_0} =0, \quad 
    \Delta \rho|_{m=m_0}=0
\end{align*}

Consider a small perturbation around $(\rho_0, m_0)$:
\begin{equation*}
    \rho = \rho_0 + \delta\rho, \quad m= m_0 + \delta m,
\end{equation*}
under which the refined mean-field equations becomes:
\begin{subequations} \label{eq:aim_rfmf0}
\begin{align}
    \partial_t (\rho_0+\delta \rho) &= D\Delta (\rho_0+\delta \rho) - v \partial_x (m_0+\delta m) \\
    \partial_t (m_0+\delta m) &= D\Delta (m_0+\delta m) - v \partial_x (\rho_0+\delta \rho) \\ \notag
    &+ I(\rho_0+\delta\rho,m_0+\delta m),
\end{align}
\end{subequations}
where $I$ is given in Eq~\eqref{eq:aim_interaction2}. 

Substituting the homogeneous steady state conditions into Eqs.~\eqref{eq:aim_rfmf0} yields:
\begin{subequations} \label{eq:aim_rfmf1}
\begin{align}
    \partial_t \delta \rho &= D\Delta \delta \rho - v \partial_x \delta m \\
    \partial_t \delta m &= D\Delta \delta m - v \partial_x \delta \rho + I(\rho_0+\delta\rho,m_0+\delta m),
\end{align}
\end{subequations}

In order to have linear partial differential equations, we need to linearise the interaction term around $(\rho_0, m_0)$, keeping only the first-order terms:
\begin{subequations} \label{eq:aim_rfmf2}
\begin{align}
    \partial_t \delta \rho &= D\Delta \delta \rho - v \partial_x \delta m \\
    \partial_t \delta m &= D\Delta \delta m - v \partial_x \delta \rho + \delta\rho\frac{\partial I}{\partial \rho}\bigg|_{(\rho_0, m_0)} + \delta m\frac{\partial I}{\partial m}\bigg|_{(\rho_0, m_0)}.
\end{align}
\end{subequations}
where
\begin{align*}
    \frac{\partial I}{\partial \rho}\bigg|_{(\rho_0, m_0)} &= 2 \sinh(H) + \frac{2\tilde{\gamma}}{\rho_0^2}m_0 - \frac{C_2m_0^2}{\rho_0^2}+\frac{2C_3m_0^3}{\rho_0^3}\\
    \frac{\partial I}{\partial m}\bigg|_{(\rho_0, m_0)} &=C_1(\rho_0) + 2C_2\frac{m_0}{\rho_0}+3C_3\frac{m_0^2}{\rho_0^2},
\end{align*}
where $C_1, C_2, C_3,\tilde{\gamma}$ are defined in Eqs~\eqref{eq:aim_interaction2}.

Using Fourier stability method, we let:
\begin{align*}
    \delta \rho &=\hat{\rho}\exp(\lambda t + iqx)\\
    \delta m &=\hat{m}\exp(\lambda t + iqx)
\end{align*}
where $q$ is the wave number. Substituting into Eqs~\eqref{eq:aim_rfmf2} yields:
\begin{equation}
    \lambda\begin{bmatrix} \hat{\rho} \\ \hat{m} \end{bmatrix}= \underbrace{\begin{bmatrix} D(-||q||^2) & -v(iq) \\ -v(iq)+\partial_\rho I & D(-||q||^2)+\partial_m I \end{bmatrix}}_{M(q)}
    \begin{bmatrix} \hat{\rho} \\ \hat{m} \end{bmatrix},
\end{equation}
where $\lambda$ is the eigenvalue of $M(q)$. 

The homogeneous steady state solutions $(\rho_0, m_0)$ are linearly stable if for all possible wave number $q$, eigenvalues of $M(q)$ satisfy:
\begin{equation*}
    \Re(\lambda)<0.
\end{equation*}

Since diffusion often stabilises for large $q$ (short wavelength), we test all $0\leq q \leq 10$ with increments 0.0075. The eigenvalues $\lambda$ are solved numerically for each $M(q)$ using Python package \texttt{numpy.linalg.eigvals}.

\section{Simulation of active Ising model}\label{ap:AIM_simulation}

For the active Ising model, we use a $100 \times 100$ lattice for results in the main text and consider $50 \times 50$ and $200 \times 200$ for the finite-size analysis. We assume periodic boundary conditions for all cases. A discrete time step 
\begin{equation*}
    \Delta t = \frac{1}{4D + \exp[\beta(J + h)]} 
\end{equation*} 
is chosen to minimise the waiting time and ensures that the total transition probabilities are bounded by one. $\Delta t$ corresponds to the physical time for one lattice sweep (or $N = L_x \times L_y$ Monte Carlo steps). At each Monte Carlo step, a particle is selected to either flip its spin, hop or do nothing. For a particle with spin $\sigma$, the corresponding probabilities are: 
\begin{gather*}
    P_{\text{flip}} = \exp[-\beta\sigma (J\frac{m}{\rho} + h)]\,\Delta t,\\
    P_{\text{hop} \rightarrow} = D(1 + \sigma\epsilon)\,\Delta t, \quad P_{\text{hop} \leftarrow} = D(1 - \sigma\epsilon)\,\Delta t, \\
    P_{\text{hop} \uparrow} = P_{\text{hop} \downarrow} = D\Delta t,
\end{gather*}
and probability of no update is $1-P_{\text{flip}}-P_{\text{hop}}$, where $P_{\text{hop}}$ is the total probability of hopping. The relaxation time ranges from $10^{6}$ to $10^7$ sweeps depending on the lattice size and the choice of parameters. We run 48 independent simulations for each control parameter combination. For each simulation, $5000$ samples are taken after the system is fully relaxed, with one sample per sweep. The final quantities are computed by averaging over all simulations. The range of parameter values we explored is summarised in table~\ref{tab:aim_parameters}.

\begin{table}[b]
    \caption{\label{tab:aim_parameters} Range of hyperparameter and control parameter values explored for the active Ising model.}
    \begin{ruledtabular}
    \begin{tabular}{ccc}
    \textrm{Parameter}&
    \textrm{}&
    \textrm{Values explored}\\
    \colrule
    Average density & $\rho_0$ &  2, 3, 5\\
    Self-propulsion speed & $\epsilon$ & 0.3, 0.5, 0.9\\
    Inverse temperature & $\beta$ & 1\\
    Diffusion rate & $D$ & 1\\
    External magnetic field & $h$ & [0.025, 0.2] \\
    Coupling strength & $J$ & [0.5, 2.5] \\
    \end{tabular}
    \end{ruledtabular}
\end{table}

We compute entropy for each simulation by first shifting the system to the comoving frame, and then constructing the joint probability distribution $P(x, n^+, n^-)$ from all $5000$ samples. The entropy is then given by:
\begin{equation*}
    S = -\sum_{x, n^+, n^-} P(x, n^+, n^-) \ln P(x,n^+, n^-),
\end{equation*}
where the sum is taken over all possible values of $x, n^+, n^-$.

Given configuration $\xi=\{m_1,\dots,m_N, \rho_1, \dots, \rho_N\}$, the work related to a small perturbation $\delta J$ is obtained from the corresponding change in the Hamiltonian \cite[Sec 3.2.2]{seifert2025Stochastic}, assuming that the system's configuration does not change during the perturbation:
\begin{align*}
    W_{J \to J+\delta J} &= \mathcal{H}(\xi; J+\delta J) - \mathcal{H}(\xi; J) \\
    &= \sum_i [ -(J+\delta J)(\frac{m_i^2}{2 \rho_i} - \frac{1}{2}) - hm_i] \\
    &\quad\quad - \sum_i [ -J(\frac{m_i^2}{2 \rho_i} - \frac{1}{2}) - hm_i] \\
    & = -\sum_i \delta J(\frac{m_i^2}{2 \rho_i} - \frac{1}{2}).
\end{align*}
We computed this quantity for each sampled configuration and averaged over samples to obtain the mean work for one simulation.

We calculate the steady state entropy production rate $\dot{S}$ by sampling all attempted particle updates (spin flips and hops) over an observation time ($\tau_{\mathrm{obs}}$), and estimate $\dot{S}$ using:
\begin{equation*}
    \hat{\dot S} = \frac{1}{\tau_{\mathrm{obs}}} \sum_{n} \ln\frac{W(\xi_n\to \xi_{n+1})}{W(\xi_{n+1}\to \xi_n)},
\end{equation*}
where $\xi_n$ is the microscopic configuration before the n-th accepted update, and $W(\xi_n\to \xi_{n+1})$ and $W(\xi_{n+1}\to \xi_n)$ are the corresponding forward and reverse transition rates. The sum is taken over all accepted updates. Rejected moves and unbiased vertical hops contribute zero to the numerator but still advance time. Total observation time is given by $\tau_{\mathrm{obs}} =$ number of sweeps sampled $\times \Delta t$, where $\Delta t$ is the discrete time step defined above.

\begin{figure}[h]
     \centering
     \begin{subfigure}[b]{0.5\textwidth}
         \centering
         \includegraphics[width=\textwidth]{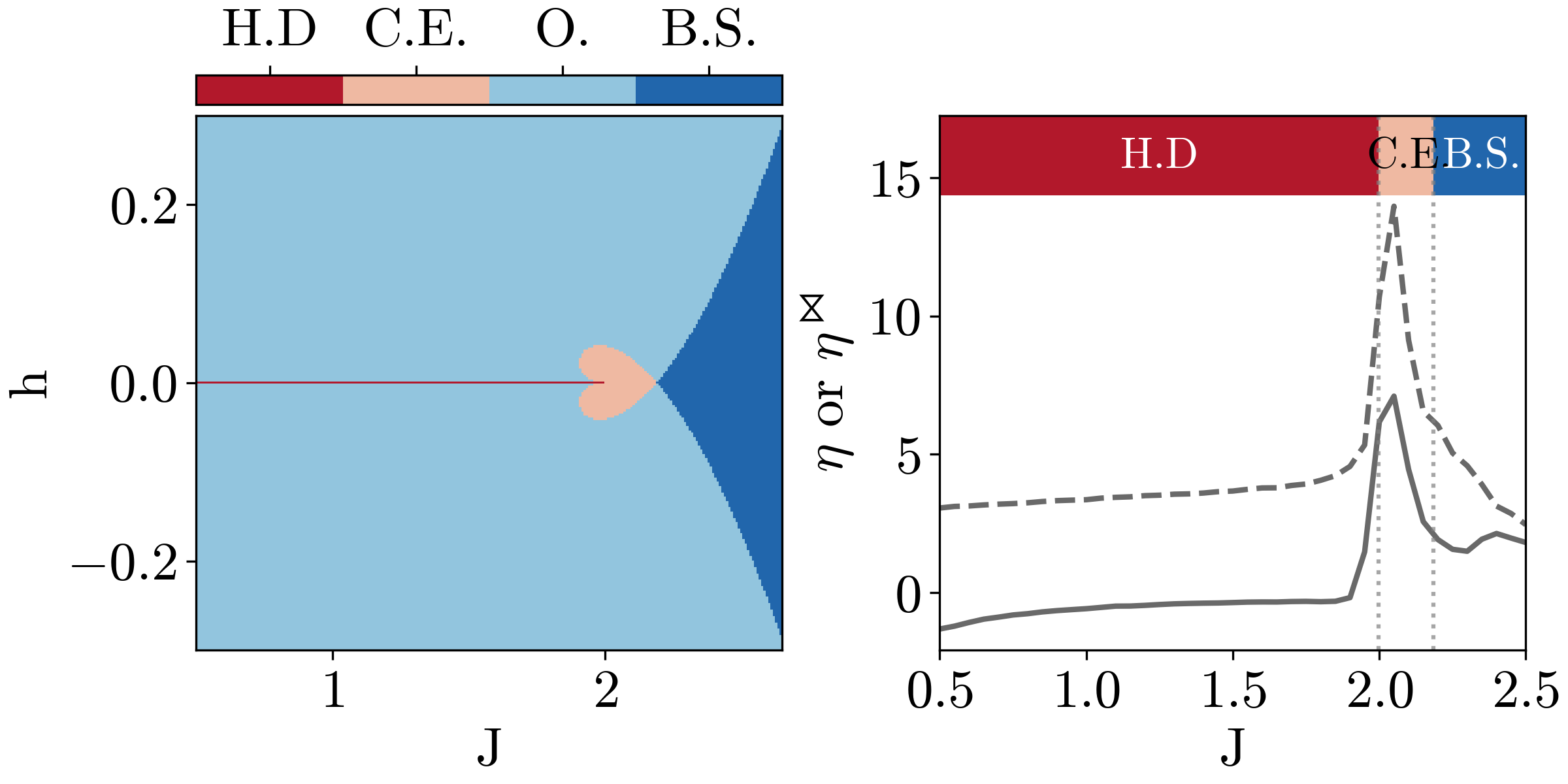}
         \caption{$\rho_0=2, \epsilon=0.3$}
     \end{subfigure}
     \vfill
     \begin{subfigure}[b]{0.5\textwidth}
         \centering
         \includegraphics[width=\textwidth]{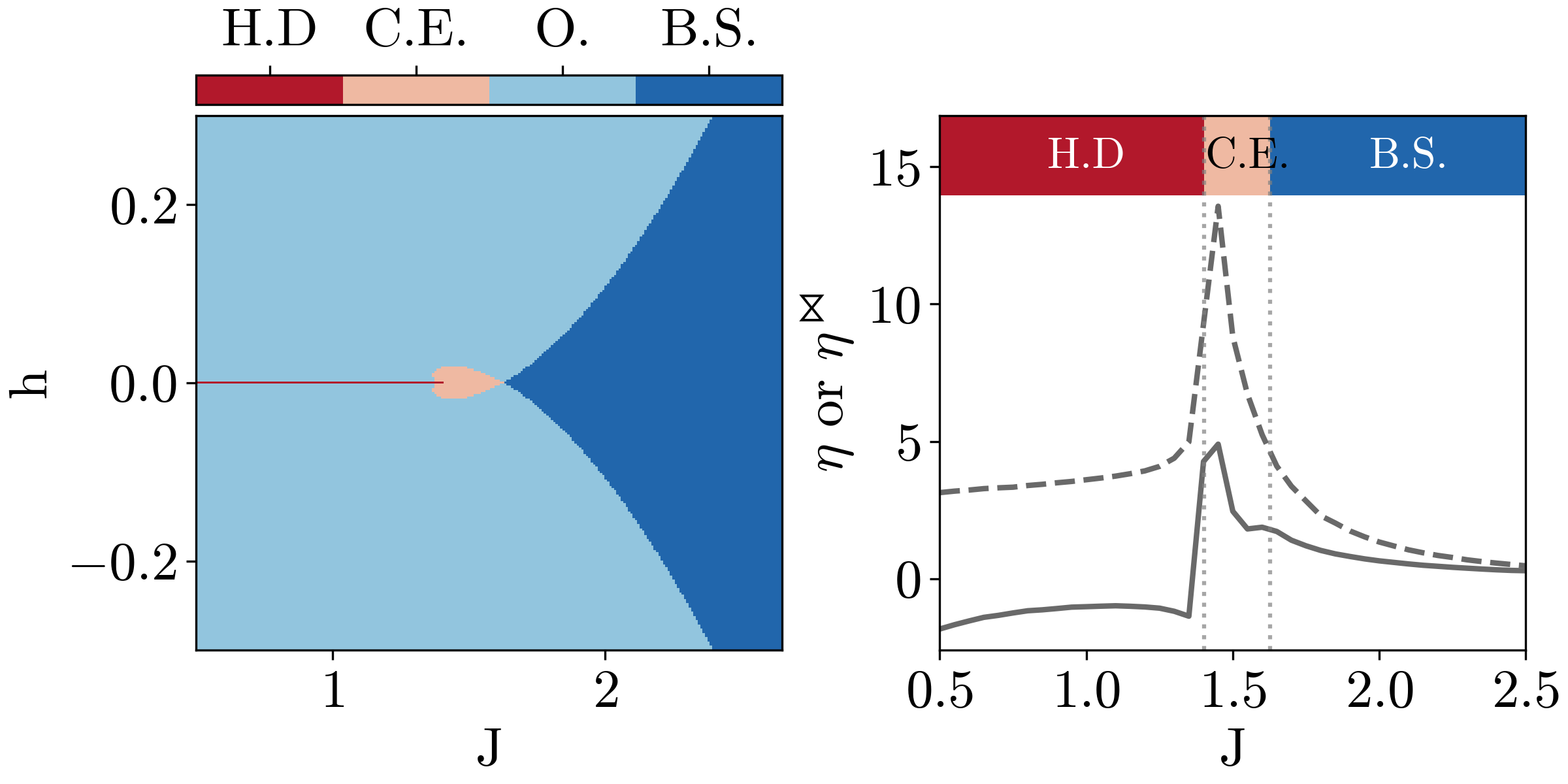}
         \caption{$\rho_0=5, \epsilon=0.5$}
     \end{subfigure}
     \caption{\textbf{Phase diagram and efficiencies for different hyperparameters}. \textbf{Left}: phase diagrams for external magnetic field $h$ and coupling strength $J$. \textbf{Right}: thermodynamic efficiency (solid curve) and inferential efficiency (dashed curve) plotted against $J$. Dotted lines indicate the two phase transition points. H.D. = homogeneous disorder, C.E. = coexistence, O. = ordered, B.S. = bistable. For both cases, lattice size = $50 \times 50, \beta=1, h=0.0$.}
     \label{fig:AIM_hyperparameter}
\end{figure}

To check robustness of the results, we also performed simulations with different hyperparameters. We found that the qualitative behaviour of $\eta(J)$ and $\etaInf(J)$ is robust to changes in $\rho_0, \epsilon$. The critical point $J_c$ shifts with different hyperparameters, but the peak of $\eta(J)$ and $\etaInf(J)$ always occurs near the phase transition. Figure \ref{fig:AIM_hyperparameter} shows the results for $\rho_0=2, \epsilon=0.3$ and $\rho_0=5, \epsilon=0.5$. Note that signature for the second transition from coexistence phase to homogeneous ordered phase is more obvious for $\rho_0=2, \epsilon=0.3$ than for $\rho_0=5, \epsilon=0.5$.

\begin{figure}[h]
    \centering
    \includegraphics[width=0.48\textwidth]{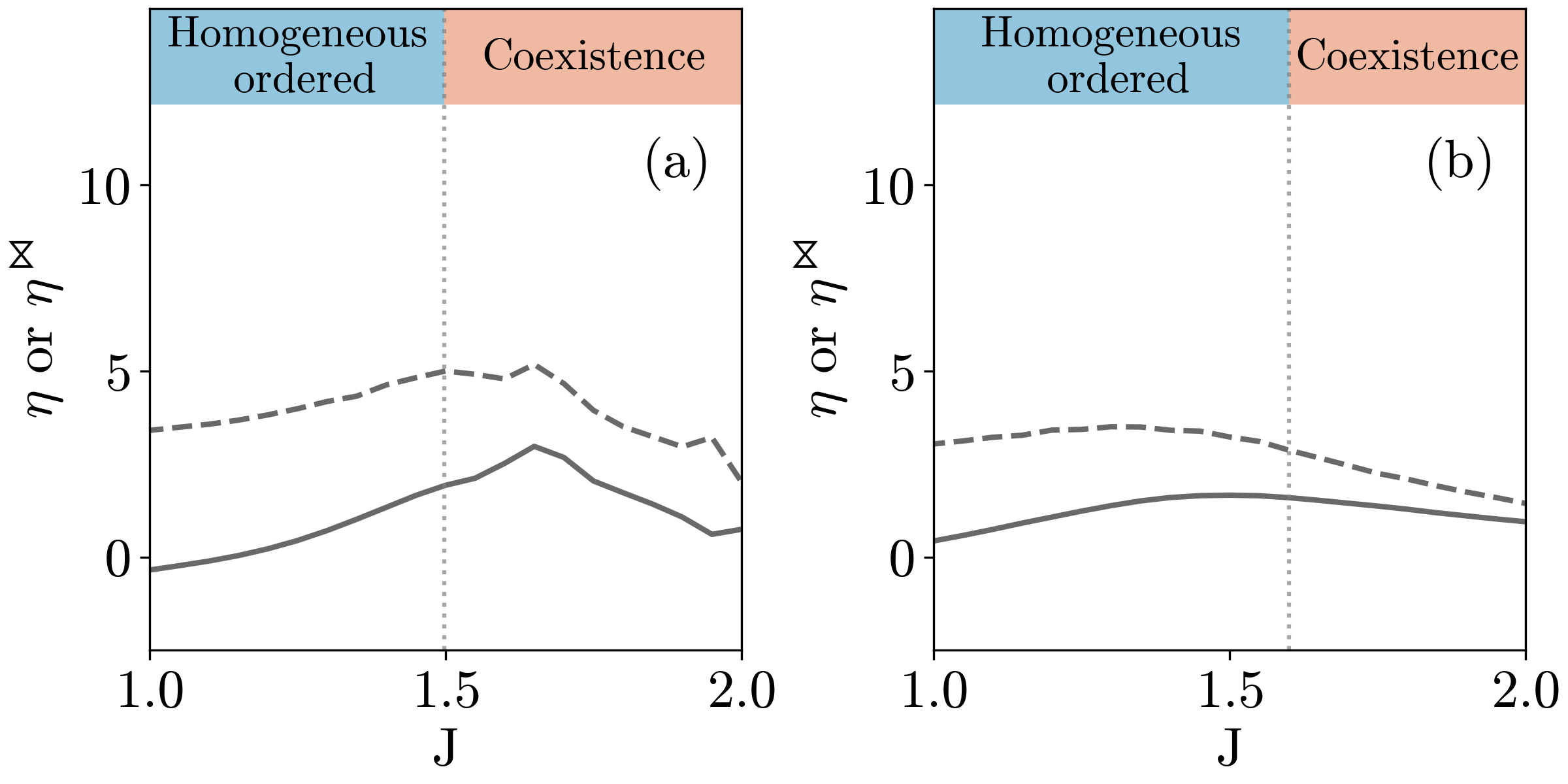} 
    \caption{\textbf{Efficiencies of active Ising model at different external magnetic fields.} Thermodynamic efficiency $\eta$ (solid curve) and inferential efficiency $\etaInf$ (dashed curve) are plotted against control parameter $J$. Vertical dotted line indicates the phase transition point predicted by the refined mean-field model. \textbf{(a)} $h=0.05, J_c \approx1.4976$; \textbf{(b)} $h=0.10, J_c \approx1.6008$. Lattice size $=100\times 100, \rho_0=3, \beta=1, \epsilon=0.9$.}
    \label{fig:AIM_eta_nonzero_h}
\end{figure}

Complementing Figure~\ref{fig:AIM_eta_zero_field}, Figure~\ref{fig:AIM_eta_nonzero_h} shows the thermodynamic efficiency ($\eta$) and inferential efficiency ($\etaInf$) for non-zero external magnetic fields $h=0.05$ and $h=0.1$. In both cases, $\eta(J)$ and inferential efficiency $\etaInf(J)$ peak near the phase transition predicted by the refined mean-field model, and $\etaInf(J)$ is consistently higher than $\eta(J)$, reflecting the nonequilibrium nature of the system of active particles.

\begin{figure}[h]
     \centering
     \begin{subfigure}[b]{0.13\textwidth}
         \centering
         \includegraphics[width=\textwidth]{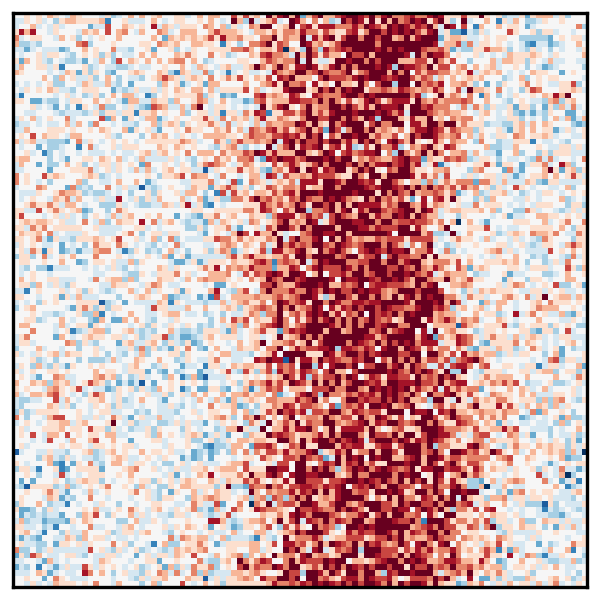}
         \caption{$100\times 100$}
     \end{subfigure}
     \hspace{0.5em}
     \begin{subfigure}[b]{0.26\textwidth}
         \centering
         \includegraphics[width=\textwidth]{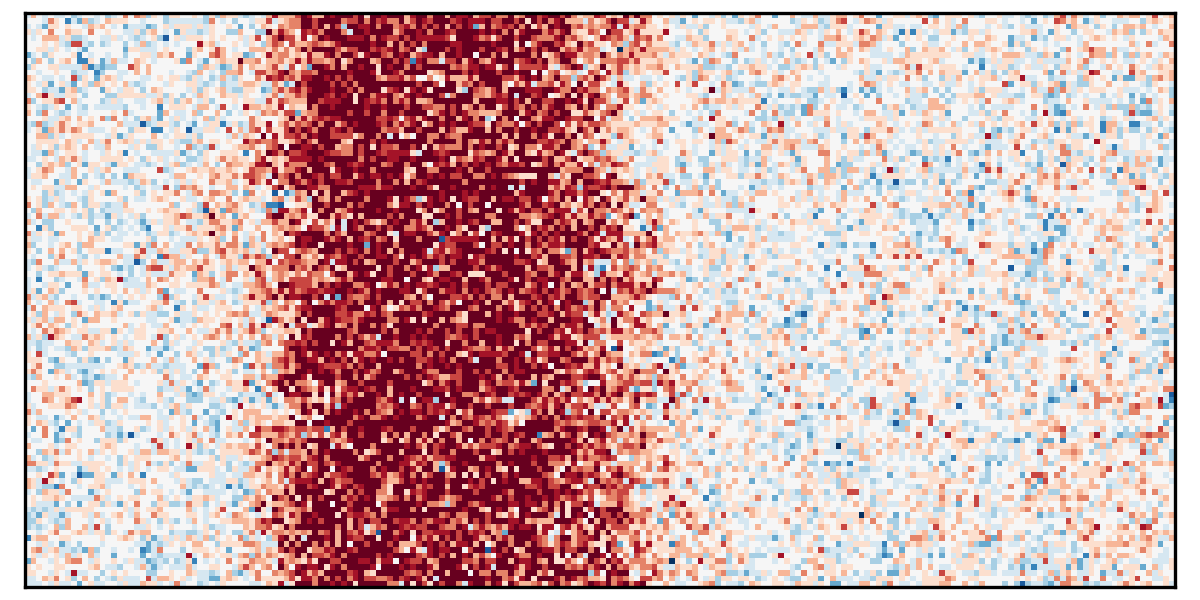}
         \caption{$200 \times 100$}
     \end{subfigure}
     \vspace{0.5em} 
     \begin{subfigure}[b]{0.13\textwidth}
         \centering
         \includegraphics[width=\textwidth]{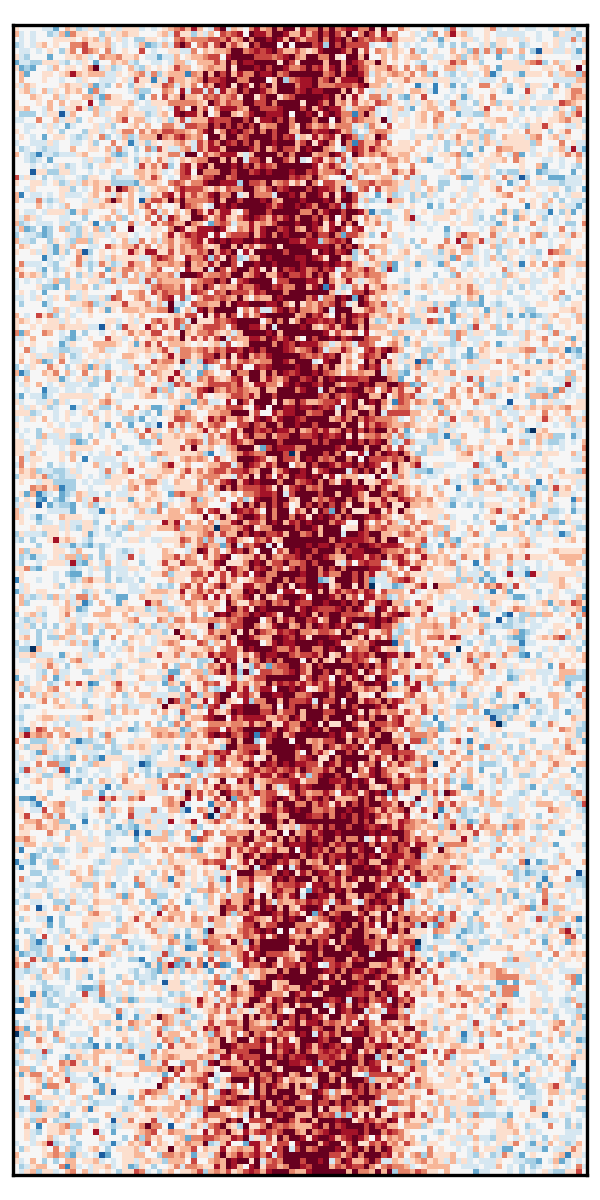}
         \caption{$100\times 200$}
     \end{subfigure}
     \hspace{0.5em}
     \begin{subfigure}[b]{0.26\textwidth}
         \centering
         \includegraphics[width=\textwidth]{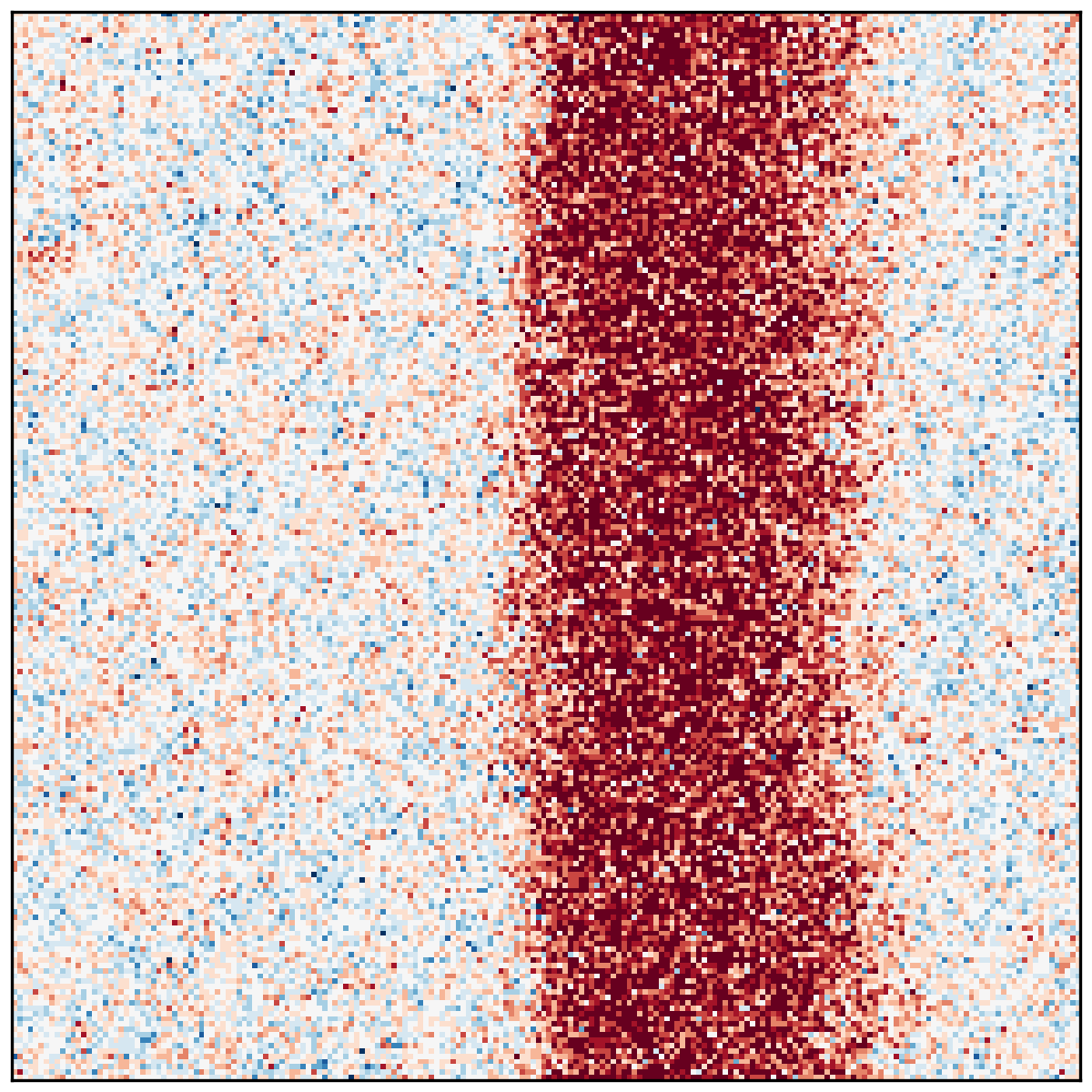}
         \caption{$200 \times 200$}
     \end{subfigure}
     \caption{\textbf{Snapshots of steady state systems in the coexistence phase, for different lattice sizes}. $\rho_0=3, \beta=1, \epsilon=1$ for all lattice sizes.}
     \label{fig:aim_finite_size_snapshots}
\end{figure}

We perform finite-size analysis for lattice sizes $50\times 50, 100\times 100$ and $200 \times 200$. Across different lattice sizes, we observe similar trends and consistent behaviour near the phase transitions from homogenous disorder phase to coexistence phase. The inferential efficiency with $50\times 50$ lattice size shows a higher peak than $100\times 100$ and $200\times 200$ due to a simulation artefact from the discretisation: the exact points of sharp changes in the numerator and the denominator are misaligned by one $\delta J$. Figure~\ref{fig:aim_finite_size_snapshots} shows snapshots of systems in NESS of different lattice sizes.

\begin{figure}[h]
    \centering
    \includegraphics[width=0.48\textwidth]{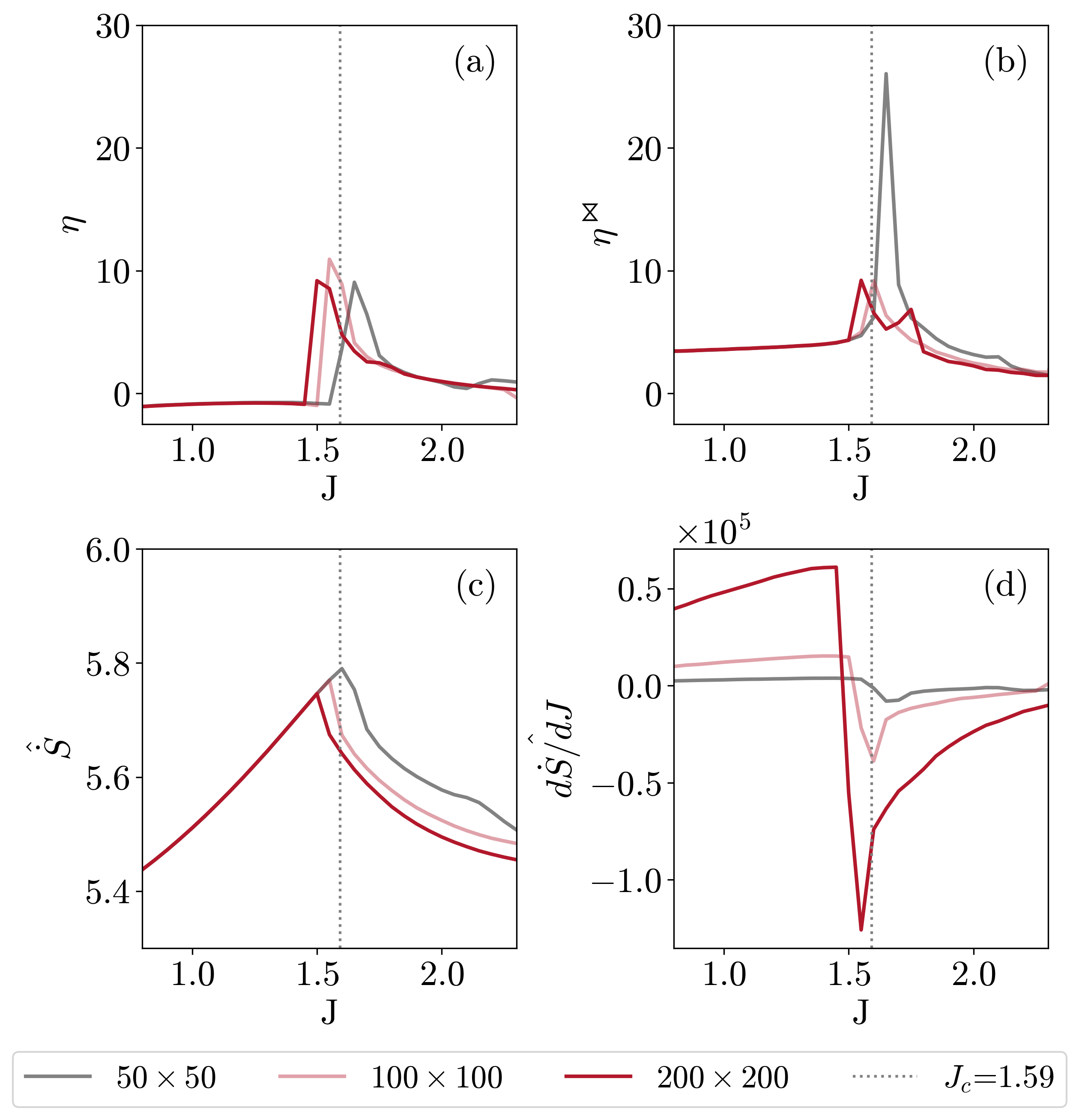}
    \caption{\textbf{Finite-size analysis.} Lattice sizes $50\times 50, 100\times 100$ and $200 \times 200$ are compared. \textbf{(a)} Thermodynamic efficiency $\eta(J)$. \textbf{(b)} Inferential efficiency $\etaInf(J)$. \textbf{(c)} Entropy production rate $\dot{S}$. \textbf{(d)} Derivative of entropy production rate $d\dot{S}/dJ$. Dotted line indicates $J_c \approx 1.5922$. For all lattice sizes, $\rho_0=3, \beta=1, \epsilon=0.9$.}
    \label{fig:AIM_finitesize}
\end{figure}

\clearpage

%

\end{document}